\documentclass[utf8]{frontiersFPHY} 

\usepackage{url,hyperref,lineno,microtype,subcaption}
\usepackage[onehalfspacing]{setspace}
\usepackage{aas_macros}         
\usepackage{textcomp,gensymb}   
\usepackage{bm}                 
\usepackage{graphicx}           
\graphicspath{{./figures/}}     
\usepackage[normalem]{ulem}

\setcitestyle{square} 

\newcommand{\Ercm}[1]{\ensuremath{E_{r}^{\text{c.m.}} = #1}~keV}
\newcommand{\nap}{$^{13}$N($\alpha$,p)$^{16}$O}
\newcommand{\clt}{$^{13}$C($^{7}$Li,t)$^{17}$O}

\def\keyFont{\fontsize{8}{11}\helveticabold }
\def\firstAuthorLast{} 
\def\Authors{Fa\"irouz Hammache\,$^{1,*}$ and Nicolas de S\'er\'eville\,$^{1,*}$}
\def\Address{$^{1}$ Universit\'e Paris-Saclay, CNRS/IN2P3, IJCLab, 91405 Orsay, France}
\def\corrAuthor{Fa\"irouz Hammache}
\def\corrEmail{fairouz.hammache@ijclab.in2p3.fr}

\begin{document}
\onecolumn
\firstpage{1}

\title[Transfer reactions in nuclear astrophysics]{Transfer reactions as a tool in Nuclear Astrophysics} 

\author[\firstAuthorLast ]{\Authors}    
\address{}                              
\correspondance{}                       

\extraAuth{Nicolas de S\'er\'eville \\  nicolas.de-sereville@ijclab.in2p3.fr}
\maketitle

\begin{abstract}
Nuclear reaction rates are one of the most important ingredients in describing how stars evolve. The study of the nuclear reactions involved in different astrophysical sites is thus mandatory to address most questions in nuclear astrophysics. Direct measurements of the cross-sections at stellar energies are very challenging - if at all possible. This is essentially due to the very low cross-sections of the reactions of interest (especially when it involves charged particles), and/or to the radioactive nature of many key nuclei. In order to overcome these difficulties, various indirect methods such as the transfer reaction method at energies above or near the Coulomb barrier are used to measure the spectroscopic properties of the involved compound nucleus that are needed to calculate cross-sections or reaction rates of astrophysical interest. In this review, the basic features of the transfer reaction method and the theoretical concept behind are first discussed, then the method is illustrated with recent performed experimental studies of key reactions in nuclear astrophysics.

\tiny
 \keyFont{ \section{Keywords:} transfer reactions, angular distributions, DWBA, spectroscopic factors, nuclear astrophysics} 
\end{abstract}

\section{Introduction}
Our understanding of stellar evolution in the Universe has been largely improved thanks to the interaction between three fields: observation, stellar modeling and nuclear physics.  All these fields are in constant development: new telescopes and satellites open more and more windows on the cosmos, stellar modeling relies on ever-increasing computing and nuclear physics takes advantage of new facilities (radioactive beams, high-intensity beams, underground laboratories) and sophisticated detection systems.
 
Nuclear reaction rates are one of the most important ingredients in describing how stars evolve. The study of the nuclear reactions involved in different astrophysical sites is thus essential
to address most questions in nuclear astrophysics. 

Experimental techniques for determining cross sections fall into two main categories: direct measurements, in which the reaction of interest is reproduced, 
even though the energy range may be different from that of the stellar site
and indirect measurements, in which a different reaction is coupled with theoretical modeling to 
obtain the cross-section of interest or to access the spectroscopic properties (excitation energies, spins and parities, decay widths,...) of the nuclei involved. 

Direct measurements at stellar energies are very challenging - if at all possible. 
This is mainly due to the very small cross-sections (sub nanobarns) of the reactions of interest (in particular when charged particles are involved), and/or to the radioactive nature of many key nuclei.

Although direct measurements of the charged-particle cross-sections are possible at the energies of interest in some cases, they are often carried out at higher energies and then extrapolated down to the energies of astrophysical interest using $\mathcal{R}$-matrix calculations for example~\cite{Lane1958}. However,  these extrapolations can easily lead to erroneous results; for example if they do not take into account the contribution of possible unobserved resonances at very low energies, or if they neglect the contribution of sub-threshold resonances.The effect of these resonances may change the extrapolated cross-section at astrophysical energies by a tremendous factor (sometimes orders of magnitude). 

The other issue concerning direct measurements is due to the radioactive nature of the nuclei involved in the reactions occurring in explosive sites (classic novae, supernovae, X-ray bursts, ...) or in the radiative captures (n,$\gamma$) in the $r$-process \cite{THIE11,ARN07,MUMP15} and sometimes in the $s$-process \cite{KAPP11}. Here, the cross sections at stellar energies are often substantial but their study requires either the production of radioactive beams (which intensity
is often weak, rarely exceeding 10$^5$ or 10$^6$ pps) or, for nuclei with 
relatively long half life, the production of radioactive targets with a sufficiently large areal density, which is often very difficult.
Therefore, direct measurements of such reactions are very challenging, and in the case of r-process reactions, they are impossible.

To overcome these problems (sub--threshold resonances, radioactive nuclei,...) indirect techniques such as transfer reaction method~\cite{Sat83}, Coulomb dissociation method~\cite{{Bert85},{Baur86},{Baur94}}, Asymptotic Normalization Coefficient (ANC) method~\cite{{MUKH90},{Mukh99},{Mukh01}}, surrogate reactions~\cite{Ratkiewicz2019} and Trojan Horse Method (THM)~\cite{{Baur1986},{TYP03},{Spitaleri2011}} are good alternatives. In these various methods, the experiments are usually 
carried out at higher energies than the Coulomb barrier which implies higher cross-sections than in direct measurements. Moreover these methods allow also 
the use of stable beams to study reactions involving  radioactive nuclei not far from the valley of stability. 
However these methods lead to results which depend on the choice of the model and its parameters in addition to the experimental errors. This is why, to reduce the overall uncertainty on the cross sections of the reactions, it is important to combine various experimental approaches.

We would like to emphasize that ANC and Trojan Horse methods as well as surrogate method are also based on transfer reactions. However, the THM and ANC method require particular kinematics conditions. For instance the transfer reactions used in ANC method need to be performed at energies where the reaction process is very peripheral in order to deduce an ANC value weakly sensitive on the potential parameters. Concerning the Trojan Horse Method, it consists in obtaining information on the two-body reaction of astrophysical interest at low relative energies by studying a three body reaction at energies above the Coulomb barrier. The basic idea of this method relies on the assumption that the three body reaction can occur via a quasi free reaction mechanism that is dominant at particular energies and angles.
For the surrogate method, the transfer reaction is used to populate the resonant states of interest and then measure their decay probability to deduce the cross section of the reaction of interest from the product of the measured quantity and the calculated compound nucleus formation cross-section. All these particular transfer reactions will not be discussed in this manuscript except the ANC method which will be described a little bit more in section ~\ref{sec:ANC}.

In this review we focus on the transfer reaction method where a composite nucleus is produced in a two body reaction by transferring one or several nucleons from a projectile to a target nucleus. Transfer reactions are a unique tool to access key spectroscopic information concerning the structure of the composite nucleus. In particular the spectroscopic factor ($C^2S$), which is related to the overlap between the wave function of the composite nucleus configuration with the one of the target nucleus, is a prime objective of such studies.

In the next section we will present the type of reactions (resonant and direct capture) which can be studied with transfer reactions. In Section~\ref{sec:transfer}, the description of the method and basic theoretical concepts behind such reactions are recalled. Some examples of recently performed experimental studies using stable and radioactive beams together with a variety of detection systems are presented in Section~\ref{sec:experiments}. We then conclude with some perspectives in a last section.

\section{Nuclear reactions of astrophysical interest} \label{sec:astro}
\begin{figure}[!htpb]
  \includegraphics[clip, trim=0cm 0cm 0cm 5cm, width=\textwidth]{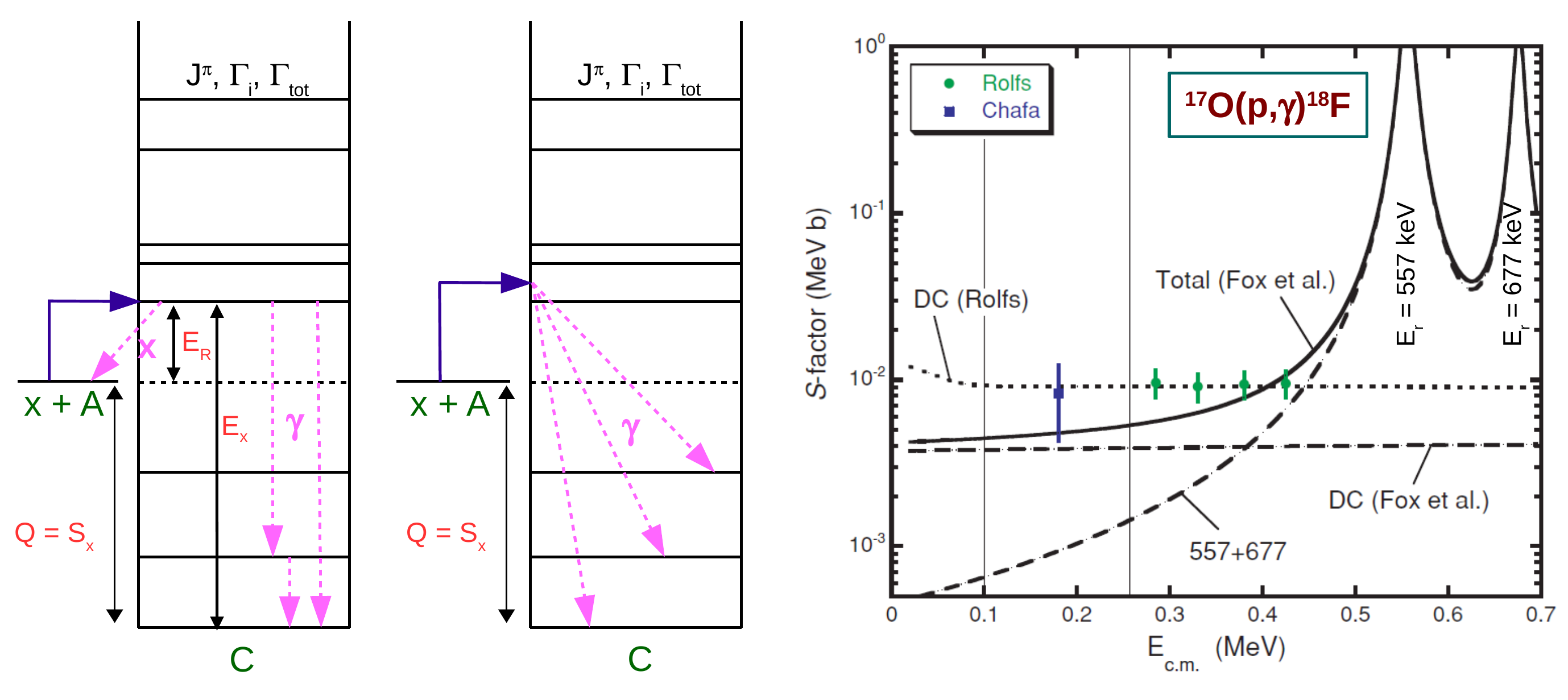}
  \caption{\label{fig:xsec}
     (Color online) Schematic view of a resonant (left panel) and direct (middle panel) capture process for a radiative capture reaction. The contribution of these two processes is shown in the case of the $^{17}$O(p,$\gamma$)$^{18}$F reaction (right panel) where the energy dependence of the astrophysical $S$-factor is presented (adapted from Ref.~\cite{Newton2010}).}
\end{figure}

Thermonuclear reaction rates are key physical inputs to computational stellar models, and they are defined per particle pair (in cm$^3$ s$^{-1}$) as~\cite{Iliadis2007}:
\begin{equation}
    \langle\sigma v\rangle = \sqrt{\frac{8}{\pi\mu}} \frac{1}{(kT)^{3/2}} \int_{0}^{\infty}\sigma(E) E e^{-E/kT}dE,
\end{equation}
where $\mu$ is the reduced mass of the interacting nuclei, $k$ is the Boltzmann constant, $T$ is the temperature at which the reaction rate is evaluated, and $\sigma(E)$ is the energy-dependent cross-section of the reaction. In order to evaluate the cross-section of nuclear reactions of astrophysical interest two processes should be considered: the resonant capture and the direct capture. Both processes are represented schematically in Figure~\ref{fig:xsec} (left and middle panel, respectively) for a radiative capture reaction $A(x,\gamma)C$, and illustrated with the case of the $^{17}$O(p,$\gamma$)$^{18}$F reaction~\cite{Newton2010} (right panel). In case of the resonant capture process (left panel) the relative energy in the center of mass between the projectile $x$ and the target $A$ must be close to the resonant energy $E_r$; the astrophysical $S$-factor\footnote{$S(E)=\sigma(E)\times E\times e^{2\pi\eta}$, where $\eta$ is the Sommerfeld parameter.} exhibits a strong energy dependence as it can be seen from the $E_r^{c.m.}=557$ and 677~keV resonances in the $^{17}$O(p,$\gamma$)$^{18}$F reaction. On the contrary the direct capture process may occur at any center of mass energy (middle panel), and the energy dependence of the $S$-factor is very smooth (see horizontal dashed lines in right panel).  Both the direct and resonant capture will now be presented in more details emphasizing the link with transfer reactions.

\vspace*{0.5 cm}
\subsection{Resonant capture} \label{sec:res}
In a resonant capture reaction $x+A \rightarrow C^* \rightarrow y +B$ the two participants of the entrance channel form an excited state of the compound nucleus $C$ which further decays into the $y+B$ exit channel. The $y$ participant can be any kind of neutral or charged particle or an electromagnetic radiation; in the latter case the reaction is called a radiative capture and $B$ is the ground state of the compound nucleus. As represented in Figure~\ref{fig:xsec} (left panel) the resonant capture only occurs at relative energies in the center of mass very close to the resonance energy defined as $E_r = E_x - Q$, where $E_x$ is the energy of the excited state in the compound nucleus, and $Q$ is the $Q$-value of the radiative capture reaction. The cross-section of a resonance is conveniently described by the one-level Breit--Wigner formula~\cite{Iliadis2007}:
\begin{equation}
    \label{eq:bw}
    \sigma(E)= \frac{\lambda^2}{4\pi}  \frac{2J_{C^*}+1}{(2J_A+1)(2J_x+1)}\frac{\Gamma_x\Gamma_y}{(E-E_r)^2+\Gamma^2/4},
\end{equation}
where $\lambda$ is the de Broglie wavelength, $J_A$, $J_x$ and $J_{C^*}$ are the spin of the entrance channel participants and of the excited state in the compound nucleus, respectively. The partial widths $\Gamma_x$ and $\Gamma_y$ represent the probability of formation and decay of the compound nucleus in its excited state $E_x$, respectively. The total width of the excited state is given by $\Gamma=\Gamma_x+\Gamma_y+ ...$ Because of the time reverse invariance of the electromagnetic and nuclear processes, the probability of formation of a given state in the compound nucleus or its decay, from or to the $x+A$ channel, is characterised by the same particle width $\Gamma_x$.

Transfer reactions are powerful tools to derive several quantities needed to calculate the cross-section given by Equation~\ref{eq:bw}. They can be used to determine the transferred orbital angular momentum $\ell$, which $(i)$ allows the determination of the parity of the compound nucleus state, and $(ii)$ may help to constrain its spin. Transfer reactions are also used to determine the excitation energies (and therefore resonance energies for unbound states), and the partial widths of the compound nucleus states~\cite{Bardayan2016}. The partial width associated to the formation of the compound nucleus in a given excitation energy ($C^*$) is given by the product of the spectroscopic factor and the single-particle width:
\begin{equation}
    \Gamma_x = C^2S_x \times \Gamma^{s.p.}_x. 
\end{equation}

The single-particle width is the decay probability of the compound nucleus state ($C^*$) when it is considered as a pure $x+A$ (particle-core) configuration. However, nuclear states are in most of the cases an admixture of configurations, and the spectroscopic factor $C^2S_x$ is related to the overlap probability between the antisymmetrized wave function of the $x+A$ channel and the compound nucleus state $C^*$. The spectroscopic factor is one of the quantities derived from the analysis of transfer reactions.

\vspace*{0.3 cm}
\subsection{Direct capture} \label{sec:direct}
The direct capture is an electromagnetic process which can not be neglected at low energies, and which may even be dominant for radiative captures where the level density is low~\cite{Mohr09} and the compound states lie at higher energies than the energies of interest~\cite{Hamm10}.

For $A(x,\gamma)C$ capture reactions where the direct capture component is dominant, the capture occurs on bound states of the final nucleus in a one step process (see Figure~\ref{fig:xsec}, middle panel). The direct capture is possible at all bombarding energies and the cross-section varies smoothly with the energy. The total cross-section of the direct capture process is given by the following expression \cite{Goriely97}: 
\begin{equation}
    \sigma_{total}^{DC}(E)=\sum_iC^2_iS_i\sigma_i^{DC}(E),
\end{equation}
where $E$ is the energy of the incident projectile and the sum runs over all available final bound states $i$ of the residual nucleus, $C_i^2S_i$ is the spectroscopic factor of the final state $i$ (see Section~\ref{sec:gen}), and $\sigma_i^{DC}$ is the calculated DC cross-section described by the following transition matrix element \cite{Kraus96}:  

\begin{equation}
\sigma_i^{DC} \propto \frac{1}{2I_A+1}\frac{1}{2S_x+1} \int d\Omega\sum_{M_AM_xM_C\sigma}\left |T_{M_AM_xM_C,\sigma} \right |^2,
\end{equation}
where $I_A$, $I_C$ and $S_x$ ($M_A$, $M_x$ and $M_C$) are the spins (magnetic quantum numbers) of the target nucleus $A$, residual nucleus $C$ and projectile $x$ respectively and $\sigma$ the polarization of the electromagnetic radiation, it can be $\pm$1.

In case of a dipole transition, $T=T^{E1}$. This transition depends on the overlap integrals of the radial parts of the bound-state wave function in the exit channel $u_{l_bI_C}(r)$, the scattering wave function $\chi_{l_x,j_x}$ in the entrance channel and the transition operator $O^{E1}$ \cite{Kim81}
\begin{equation}
 I^{E1}_{l_bI_C;l_x,j_x}  \propto \int dr \, u_{l_bI_C}(r)O^{E1}(r)\chi_{l_xj_x}(r) ,
\end{equation}
where $l_b$ is the orbital angular momentum of $A+x$ two clusters in the nucleus $C$. The complete DC formalism can be found in Ref.~\cite{Kim81} and the computer code TEDCA~\cite{Krauss92} can be used to calculate its cross-section. 

\section{Transfer reaction method} \label{sec:transfer}
Transfer reactions in which one nucleon or a cluster of nucleons are exchanged between the target and the projectile are often used in nuclear structure studies to determine the energy position and the orbital occupation of the excited states of many nuclei. Likewise it is widely used in nuclear astrophysics to determine the partial decay widths of nuclear states involved in resonant reactions, and to evaluate the direct capture cross-section. 

\vspace{0.3 cm}
\subsection{General concepts} \label{sec:gen}
\begin{figure}[h]
\begin{center}
    \includegraphics[width=9cm]{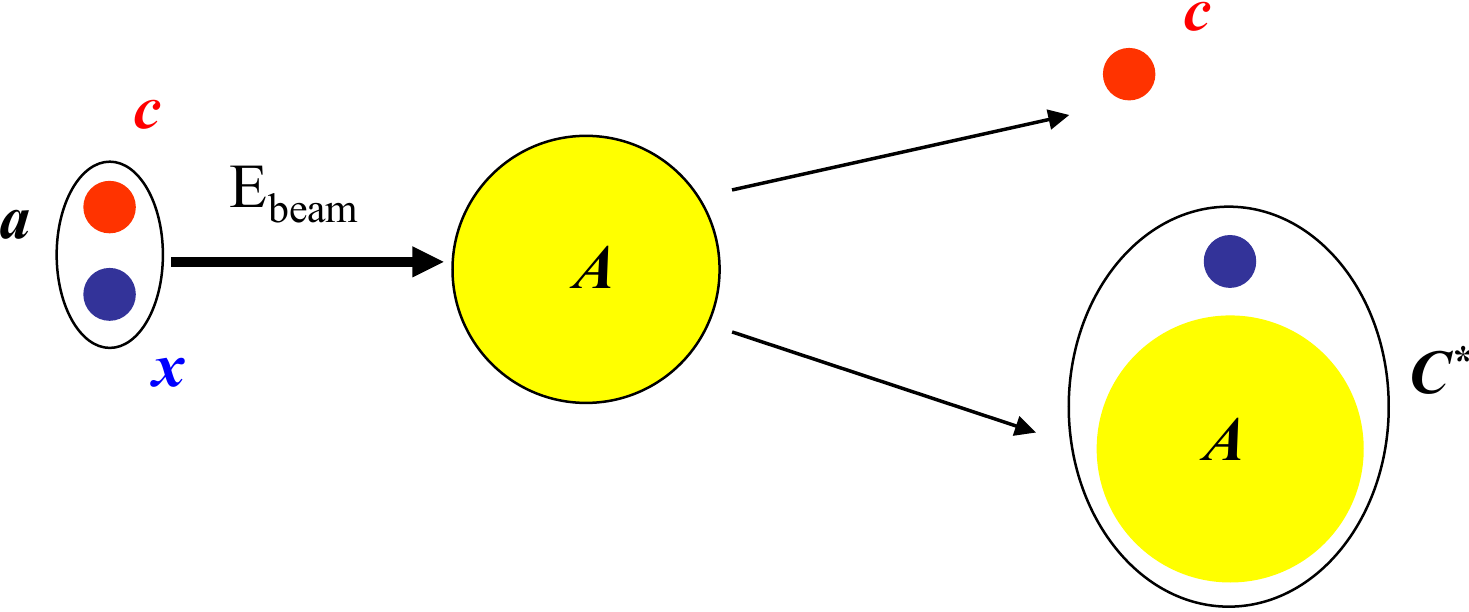}
    \caption{Sketch of a transfer reaction where the particle $x$ is transferred from the projectile $a$ to the target nucleus $A$ forming the final state $C^*$.}
    \label{f:Transfer-scheme}
\end{center}    
\end{figure}

Let's consider the simple case of a radiative capture $A(x,\gamma)C$. Whether the reaction proceeds through a resonant or direct capture, the spectroscopic factor $C^2S_x$ of the unbound or bound states, respectively, in the compound nucleus $C$ is needed to evaluate the cross-section (see Section~\ref{sec:astro}). It is then relevant to populate the excited states of the compound nucleus $C$ by transferring the particle $x$, which can be a single nucleon or a cluster of nucleons, to the target nucleus $A$ (see Figure~\ref{f:Transfer-scheme}). The transfer reaction will then be $A(a,c)C$, where $a$ is a composite system made of $x$ and $c$. The valence states of the final nucleus $C$ will be populated, and the reaction mechanism will be considered as a one step direct reaction if the reaction occurs without perturbation of the target (core) nucleus $A$ or the projectile $a$~\cite{Sat83}.

Once the particle $x$ is transferred to the target $A$, the projectile component $c$ will continue its movement and should be detected. By measuring its emission angle and energy, the energy of the populated states in nucleus $C$ can be obtained using two-body kinematic properties if the masses of the interacting nuclei are known. A precise measurement of the energies of the excited states of interest is very important to calculate accurately the resonance energies involved in the evaluation of the thermonuclear reaction rates. 

From a comparison of the shape of the measured angular distributions to those predicted by theory, it is possible to deduce the transferred angular momentum $\ell$ which indicates, for single-nucleon transfer, into which orbital the nucleon has been transferred. Indeed, the shape of measured differential cross-sections exhibits features that are sensitive to the transferred orbital momentum $\ell$ and the knowledge of the latter may constrain the spin of the populated states. Note that the spin of the populated states can be obtained from polarization measurements~\cite{Knut80}. The magnitude of the differential cross-sections are sensitive to the spectroscopic strengths of the populated states and their analysis using an adequate formalism allows the extraction of the spectroscopic factor $C^2S_x$.

When the direct transfer mechanism is dominant the measured transfer angular distributions are often analyzed using the Distorted Wave Born Approximation (DWBA) formalism (see Section~\ref{sec:DWBA}). However other reaction mechanisms such as the compound nucleus mechanism, the multi-step transfer reaction mechanism, the projectile breakup and the transfer to continuum~\cite{Sat83} can occur. The contribution of these mechanisms can be evaluated by using Hauser Feshbach calculations \cite{Haus52}, coupled reaction channel calculations (CRC), Adiabatic Distorted Wave Approximation (ADWA) (see Section~\ref{sec:ADWA}) and continuum discretized coupled channel (CDCC) calculations~\cite{Nunes11}, respectively.

The angular distributions of direct reactions display a characteristic shape which often shows a forward protuberant peak and smaller peaks at larger center-of-mass angles (see Figure~\ref{fig:dsig31P} in Section~\ref{sec:si30pg}). This is in contrast to the compound nucleus mechanism where the angular distribution shows an almost flat and symmetric shape with respect to 90$^\circ$ in the center-of-mass. Hence to be more sensitive to the direct reaction mechanism, transfer measurements need to be performed at relatively small detection angles (typically $\theta_{c.m}$~$\leqslant$~50$^\circ$). 

\vspace*{0.5 cm}
\subsection{Elements of theory}
\subsubsection{Distorted Wave Born Approximation} \label{sec:DWBA}
The most commonly used theoretical model to describe direct transfer reaction cross-sections is the Distorted Wave Born Approximation (DWBA) which relies on the following assumptions:
\begin{itemize}
    \item the entrance and exit channels processes are dominated by the elastic scattering
    \item the transfer process is weak enough to be treated as a first order perturbation
    \item the nucleon(s) transfer occurs directly between the two active channels $a+A$ and $c+C$
    \item the transferred nucleon(s) is directly deposited on the final state with no rearrangement of the core configuration
\end{itemize}

The transfer reaction cross-section is proportional to the square of the transition amplitude which in case of the DWBA model, and in the \textit{post} representation\footnote{The transition amplitudes can be given in either a \textit{post} or \textit{prior} form depending on whether it is based on the interactions in the exit or entrance channel, respectively. DWBA calculations with either form are equivalent~\cite{Sat83,Thompson2009}}, is given by~\cite{Austern1970}:
\begin{equation}
    \label{eq:tamp}
    T_{i \rightarrow f}^{DWBA} =J \int \! \! \! \! \int \chi_{f}^{(-)}(\vec{k}_f,\vec{r}_f)^* \langle c,C|V_{cC}-U_{cC}|a,A \rangle  \chi_i^{(+)}(\vec{k}_i,\vec{r}_i)d\vec{r}_i d\vec{r}_f,
\end{equation}
where $\chi_i$ and $\chi_f$ are the distorted wave functions describing the elastic scattering process in the entrance and exit channel, respectively; $\vec{k}$ and $\vec{r}$ being the wave number and the relative coordinates for the considered channel, and $J$ is the Jacobian for the transformation to these coordinates. The term $V_{cC}-U_{cC}$ describes the non-elastic scattering processes, $V_{cC}$ being the sum of all interaction between $c$ and $C$ while $U_{cC}$ being the optical potential describing the $c+C$ elastic scattering. For transfer reactions where the transferred nucleon(s) are small compared to the target, the term $V_{cC}-U_{cC}$ is often approximated by the potential $V_{cx}$\footnote{The $V_{cC}$ potential can be separated in two parts: $V_{cC}=V_{cx}+V_{cA}$, which leads to $V_{cC}-U_{cC}=V_{cx}+(V_{cA}-U_{cC})$. The no-remnant approximation is often used to neglect the $(V_{cA}-U_{cC})$ term.}, and the quantity $\langle c,C|V_{cx}|a,A \rangle$ is then the form factor of the reaction. Since the $V_{cx}$ potential only acts on the projectile the form factor can be factorized as $\langle c,C|V_{cx}|a,A \rangle = \langle C|A \rangle \langle c|V_{cx}|a \rangle$. The form factor contains all the information concerning the angular momentum selection rules and the nuclear structure. It embeds the overlap function describing the transferred nucleon or group of nucleons in the projectile $a$ and in the final bound state $C$. In the latter case the radial part of the overlap function $I^{C}_{xA}(r)$ is usually approximated by a model wave-function of the bound state $C$ as follows~\cite{Mukh99}:
\begin{equation}
    \label{eq:overlap}
     I^{C}_{xA}(r) \approx S^{1/2}_{xA}\ \varphi_{xA}(r),
\end{equation}
where $\varphi_{xA}(r)$ is the radial part of the bound state wave-function describing the relative $x+A$ motion, and $S_{xA}$ is the spectroscopic factor of the $x+A$ configuration. The wave function $\varphi_{xA}(r)$ does not contain the intrinsic wave functions of $x$ and $A$. The full relation between the form factor $\langle C|A \rangle$ and the radial overlap can be found in Ref.~\cite{MacFarlane1974} (see Eqs. 3 and 4).

The spectroscopic factor $S_{xA}$ expresses the overlap probability between the $x+A$ wave-function and the final bound-state configuration $C$. It can be extracted from the ratio of the measured differential cross-section to the one calculated by the DWBA for the relevant single-particle or cluster transfer:
\begin{equation}
\left(\frac{d\sigma}{d\Omega}\right)_{exp}=S_{xA}S_{xc}\left(\frac{d\sigma}{d\Omega}\right)_{DWBA}.
\label{q:norm}
\end{equation}
The product of the spectroscopic factors corresponding to the configuration of the $x+A$ bound state ($S_{xA}$) and of the projectile ($S_{xc}$) is involved in the previous expression. Hence, by knowing one of the spectroscopic factors it is possible to extract the other one. Therefore the light projectile in transfer reaction is usually chosen to have a strong cluster configuration, e.g. $S_{xc}\approx1$, as in the case of the $(d,p)$ reaction for example.

\vspace*{0.5 cm}
\subsubsection{Finite-range and zero-range calculations} \label{sec:ZRFR}
The calculation of the DWBA transfer reaction cross-section involves the evaluation of the transition amplitude (see Equation~\ref{eq:tamp}) which is of the form of a six-dimensional integral over the two relative coordinate variables $\vec{r}_i$ and $\vec{r}_f$. In a finite-range DWBA calculation (FR-DWBA) the integral appearing in the transition amplitude is undertaken exactly over the two radial coordinates. While computational resources nowadays allow finite-range calculations to be performed rather easily, this was not always the case and the evaluation of the six-dimensional integral required some approximations. The most common is the zero-range approximation (ZR-DWBA) which relies on the assumption that the form factor has a small range, either because it is proportional to a short range interaction, or because the internal wave-function of the projectile has a small range. The physical meaning of such approximation is that the light particle in the exit channel is emitted at the same point at which the light particle in the entrance channel is absorbed. Under this assumption the DWBA transition amplitude reduces to a three-dimensional integral which is much more tractable from a numerical point of view, and only the form factor describing the interaction of the transferred particle with the core in the final nucleus has to be considered. The integrand of the transition amplitude is proportional to the product of the projectile internal wave-function and the interaction potential between its $c$ and $x$ components, i.e. $D(\vec{r}_{cx})=V_{cx}\phi_{cx}$, and in the zero-range approximation one has:
\begin{equation}
    D(\vec{r}_{cx})=D_0 \delta(\vec{r}_c - \vec{r}_x),
\end{equation}
where $D_0$ can be calculated exactly for light systems~\cite{Austern1970,Sat83}.

The zero-range (ZR) approximation is usually a good assumption when calculating the cross-section of a direct transfer reaction induced by light projectile. In the case of the typical $(d,p)$ stripping reaction this is partially justified by the small size of the deuteron in comparison to the size of the other interacting nuclei, and by the $s$-wave nature of its dominant configuration. However, the zero range assumption is no longer valid if the projectile is not in an $s$-wave internal state, or has a very large size. For these cases, finite range DWBA calculations are mandatory in order to provide reliable theoretical cross-sections.
The ($^7$Li,t) $\alpha$-particle transfer reaction provides a good example where the $\alpha+t$ system is in a relative $p$-state, thus making the ZR assumption very poor and sometimes wrong. 

\vspace*{0.5 cm}
\subsubsection{Reduced and partial decay widths}
Once the spectroscopic factor $S_x$\footnote{In the following we define $S_x \equiv S_{xA}$ in order to simplify the notations.} of the state of interest is extracted, its reduced decay width $\gamma^2_x$ can be determined using the following formulas~\cite{Bech78}:
\begin{equation}
 \gamma_x^2 = S_{x} \times \gamma_{x,s.p.}^2,
 \label{eq:reducedWidth}
\end{equation}
where $\gamma_{x,s.p.}^2$ is the single-particle reduced width defined as:
\begin{equation}
 \gamma_{x,s.p.}^2=\frac{\hbar^2 R}{2\mu}{|{\varphi}(R)|}^2,
 \label{eq:reducedWidth_sp}
\end{equation}
where $\mu$ is the reduced mass of the $A+x$ channel, and $\varphi(R)$ is the radial part of the wave function describing the relative motion of the $A+x$ system forming the bound state of $C$ calculated at a channel radius $R$. The radial part of the wave function is normalized such as $\int_0^\infty r^2\varphi(r)dr= 1$, and the radius $R$ is chosen where $\varphi(r)$ reaches its asymptotic behaviour.

In case of unbound states the partial decay width $\Gamma_x$ is related to the reduced decay width by~\cite{Rolf88}: 
\begin{equation}
    \Gamma_x=2 \ P_l(R,E) \ \gamma_x^2,
     \label{eq:partialWidth}
\end{equation}
where $P_l(R,E)$ is the Coulomb and centrifugal barriers penetrability for relative angular momentum $l$. The penetrability factor is calculated at the energy of the resonant state for the same radius $R$ as the one used to determine the reduced decay width.

A common procedure to determine the partial decay width for an unbound state is to use the weakly-bound approximation. In this approach the radial form factor is calculated for a very weakly bound state (typical binding energies between 5 keV to 50 keV) and is further used to calculate the reduced decay width using Eq.~\ref{eq:reducedWidth}. The partial width is then obtained with Eq.~\ref{eq:partialWidth} evaluated at the energy of the resonance. It has been shown that for proton or neutron resonances having single-particle widths small compared to their resonance energy, the weakly-bound approximation gives spectroscopic information within 15\% with results obtained using an unbound form factor~\cite{Ray1975}.

\vspace*{0.5 cm}
\subsubsection{Asymptotic Normalization Coefficients} \label{sec:ANC}
As mentioned in the introduction, the ANC method is a particular case of transfer reactions. It relies on the peripheral nature of the reaction process that makes the calculations free from the geometrical parameters (radius, diffuseness) of the binding potential of the nucleus of interest and less sensitive to the entrance and exit channel potentials. The ANC method was extensively used for direct proton-capture reactions of astrophysical interest where the binding energy of the captured charged particle is low \cite{Trib05} and also for reactions where the capture occurs through loosely sub-threshold resonance states~\cite{Brun99,Avil15}. These very peripheral transfer reactions performed at sub-Coulomb energies are good tools to determine asymptotic normalization coefficients (ANCs) which are weakly sensitive to the calculations and which may be linked to the partial width of a resonance~\cite{Mukh99}. Nevertheless, ANC's can also be determined from transfer reactions performed at energies above the Coulomb barrier.

The asymptotic normalisation coefficient $C$ describes the amplitude of the tail of the radial overlap function at radii beyond the nuclear interaction radius and in case of a bound state it can be related to the spectroscopic factor using the following expression~\cite{Mukh99}: 
\begin{equation}
    C^2 = S_x \frac{R^2 \varphi^2(R)}{W_{-\eta_{xA},l+1/2}^2(2k_{xA}R)},
 \label{eq:ANC}
\end{equation}
where $W_{-\eta_{xA},l+1/2}(2k_{xA}R)$ is the Whittaker function describing the asymptotic behavior of the bound state wave function, characterized by $\eta_{xA}$ the Sommerfield parameter of the $x+A$ bound state, $l$ the relative orbital momentum and $k_{xA}$ the wave number of the $x+A$ bound state. 

\vspace*{0.5 cm}
\subsubsection{Adiabatic Distorted Wave Approximation} \label{sec:ADWA}
When one of the participants of the transfer reaction is a loosely bound system, the DWBA may not be suited to analyse the data since the breakup of this system becomes an important additional reaction channel to consider. This is the case when deuterons are involved in the transfer reaction since they can break up easily into their constituents due to their small binding energy ($BE=2.224$~MeV). The Adiabatic Distorted Wave Approximation (ADWA) was developed to take into account the breakup channel, and it was first introduced in the case of the $(d,p)$ stripping reaction~\cite{Johnson1970}. In this approximation, the effective potential including the deuteron breakup is calculated by taking into account the proton and the neutron interactions with the target nucleus, as well as a corrective term describing the proton-neutron interaction~\cite{Wal76}. Various studies have shown that using the ADWA results in a substantial improvement of the description of $(d,p)$ angular distributions for deuteron energies larger than 20 MeV~\cite{King18}. A comparison of DWBA and ADWA calculations is given in Section~\ref{sec:fe60dp} for the $^{60}$Fe(d,p)$^{61}$Fe reaction.

\vspace*{0.5 cm}
An interesting feature of the ADWA method is that its implementation is very similar to the DWBA calculations, thus any pre-existing inputs for the DWBA calculations can be easily adapted to perform ADWA calculations. The only difference is in the optical model potential parameters (see Section~\ref{sec:optmod}) describing the interaction of the deuteron with the target. While in the DWBA this potential is adjusted to reproduce the elastic scattering differential cross-section, it is no longer the case for the ADWA since it includes the treatment of the deuteron breakup. Therefore the optical potential used in ADWA will not be adapted to provide a good description of the deuteron elastic scattering, but it will give instead a better description of the transfer differential cross-section.

\vspace*{0.5 cm}
\subsection{Ingredients for a DWBA calculation}
To calculate the transfer DWBA differential cross sections, a number of computer codes are available such as FRESCO~\cite{Thom88}, DWUCK~\cite{Kunz}, TWOFNR~\cite{twofnr} and PTOLEMY~\cite{ptolemy}, to cite a few of them. They all require the same ingredients which are the distorted waves in the entrance and exit channel, and the two overlap functions which describe the relative motion of the transferred nucleon in the projectile and in the final state (see Sec.~\ref{sec:DWBA}). These ingredients are calculated using optical model and interaction potentials whose parameters are the main inputs for any DWBA code.

\vspace*{0.5 cm}
\subsubsection{Distorted waves} \label{sec:optmod}
The distorted waves are the solution of the Schr\"odinger equation for elastic scattering by an appropriate optical-model potential. This potential has usually a central (both real and imaginary parts), a spin-orbit and a Coulomb component; and its most common shape is a Woods-Saxon well. The best way to determine the potential parameters is to analyse the differential cross-section of the elastic scattering in the entrance and exit reaction channel at the same energy as the reaction under study. When elastic measurements are not available, one should use potential parameters deduced from measurements performed in the mass region close to the nuclei of interest at close incident energies. Another alternative is to use global potential parametrisations obtained by fitting a large number of elastic scattering data. The radius, diffuseness and depth of the different components of the potential usually have an energy and $Z$, $A$ dependence allowing to derive a potential parameter set adapted to the reaction under study. The most commonly used global parametrisations for protons, neutrons and deuterons are those of Perey and Perey~\cite{Pere76}, Daehnick et al.~\cite{Dae80} and Koning et al.~\cite{Kon03}.

\vspace*{0.5 cm}
\subsubsection{The $\langle C|A \rangle$ overlap function} \label{sec:ff}
The radial part of the overlap function is usually approximated by the radial part of the wave-function describing the relative motion of the transferred nucleon $x$ to the core $A$ to form the bound state $C$ (see Equation~\ref{eq:overlap}). It is obtained by solving the Schr\"odinger equation for an interaction potential usually having a Woods-Saxon form. In this procedure the depth of the real part of the volume component of the potential is adjusted to reproduce the binding energy of the bound state. 

The shape of the bound state wave-function is dictated by the orbitals to which the nucleon or group of nucleons  are transferred. In the case of a single nucleon transfer reaction, the nucleon is transferred to an orbital characterized by the usual quantum numbers $(n,l)$,  where the principal quantum number $n$ gives the number of nodes of the wave-function, and the transferred orbital angular momentum $l$ is obtained from selection rules and parity conservation. The case of multi nucleon transfer is more delicate since the transferred nucleons may be dropped on different orbitals. In this case the number of nodes $N$ of the radial wave-function is obtained using the Talmi-Moshinsky relation~\cite{Mos59}:
\begin{equation}
    (2N+L) + (2n+l) = \sum_i (2n_i+l_i),
\end{equation}
where $(n,l)$ are the intrinsic quantum numbers of the transferred cluster, and $(n_i,l_i)$ characterize the orbitals to which the individual nucleons forming the cluster are transferred. Here, $N$, $n$ and $n_i$ counts the number of nodes excluding that at zero radius. The transferred angular momentum $L$ is obtained as for the single-nucleon transfer case from selection rules and parity conservation. Note that the previous relation is strictly valid for harmonic oscillator functions and hence is only approximate for a general case. A detailed example is presented in Section~\ref{sec:n13ap} for the $^{13}$C($^7$Li,t)$^{17}$O $\alpha$-particle transfer reaction.

The radial form factor strongly depends on the radius and the diffuseness of the potential. Different realistic $(r,a)$ sets can be used and the selected ones are those giving the best description of the measured angular distributions.  

\vspace*{1.0 cm}
\subsubsection{The $\langle c|a \rangle$ overlap function}
The way the projectile is treated depends on the type of DWBA calculation. If the zero-range approximation is used, then it is enough to know the value of $D^2_0$ (see Section~\ref{sec:ZRFR}), and numerical values for typical transfer reactions can be found in the literature~\cite{Sat83}. In case a finite-range DWBA calculation is considered, the same exact procedure as for the final bound state can be used to determine the radial part of the wave-function describing the relative motion of the transferred nucleon $x$ in the projectile $a$. In case of light-ions overlaps like $\langle d|n+p \rangle$ there are better choices such as the the Reid soft-core potential~\cite{Reid1968} which gives realistic wave-functions. Note also that recent advances now provide one-nucleon spectroscopic overlaps, spectroscopic factors and ANCs in light nuclei ($A\leq7$) based on the realistic two- and three-nucleon interactions using, for example, the Green's function Monte Carlo (GFMC) method~\cite{Brida2011}. 

\vspace*{0.5 cm}
\subsection{Uncertainties on spectroscopic factors, ANCs and reduced widths} 
The uncertainty associated to the extracted spectroscopic factors depends on the accuracy of the measured differential cross-sections, and mainly on the uncertainties related to the different parameters used in the DWBA calculation. This includes the optical potential parameters used to describe the wave functions of the relative motion in the entrance and exit channels, and the geometry parameters of the potential well describing the interaction of the transferred particle with the core in the final nucleus. In case of a one-nucleon transfer reaction these uncertainties give rise to a typical uncertainty on the spectroscopic factor of about 25\%--35\%~\cite{Flav18}, which is increased to 30\%--40\% in case of an $\alpha$-particle transfer reaction~\cite{{Pel08},{Oul12}}. However, the DWBA model remains very useful and even essential for reactions that cannot be studied directly and whose uncertainty on cross sections is more than a factor two. 

Concerning the reduced widths and ANCs deduced from transfer reactions (see Equations~\ref{eq:reducedWidth}-\ref{eq:reducedWidth_sp} and Equation~\ref{eq:ANC}, respectively), their uncertainties depend, not only on the spectroscopic factors uncertainty but also on the potential parameters used to calculate the radial wave function of the relative motion between the transferred nucleon(s) and the core nucleus. Since the spectroscopic factor determination also depends on the aforementioned potential parameters its uncertainty is then correlated to the determination of the radial wave-function. It is then mandatory that the same optical potential parameters must be used in deriving the spectroscopic factor and the radial wave function to determine the reduced width and the ANC. Their associated relative uncertainties may therefore be different from the one of the spectroscopic factors.

We would like to point out that the spectroscopic factors defined here are experimental quantities subject to the uncertainties mentioned above. In  theory, they can be defined properly but there is a long discussion in recent years whether they can be considered as a well-defined observable \cite{Dug15}. A direct use of proper many-body wave functions for the structure of the nuclei in the calculation of the matrix elements would remove the problem of defining spectroscopic factors and allow better testing of nuclear structure. However, these many-body calculations are up to now possible only for light nuclei \cite{{Arai11},{Navr12},{Delt14},{Raim16}}, and not for most of the nuclei involved in the various nucleosynthesis processes studied in nuclear astrophysics. 

\vspace{1 cm}
\section{Experimental needs and challenges for transfer reaction studies} \label{sec:expneeds}
We have seen so far that the analysis of experimental angular distributions obtained from two-body transfer reactions is a unique tool to access key spectroscopic information (energy of excited states, spectroscopic factors and transferred angular momentum) concerning the composite nucleus produced by transferring one or several nucleons from a projectile to a target. A sketch of such transfer reaction is given in Fig.~\ref{f:Transfer-scheme} and in the vast majority of experimental approaches the goal is to measure the energy and angle of the emitted light particle $c$. It is then possible to determine the excitation energy of the composite nucleus by using the two-body kinematic properties of the reaction. In addition, the number of light particles detected at different angles is the main ingredient used to extract the angular distribution.

While one is usually interested in a specific transfer reaction channel characterized by the light particle $c$, many other processes ((in)elastic scattering, fusion-evaporation, etc...) produce many other kinds of particles which need to be disentangled from $c$. It is therefore a requirement for the experimental detection system to have a good particle identification capability. Moreover it is important that the resolution in the center of mass be the best as possible in order to separate the different excited states of the composite nucleus. Another need for the detection system is to cover the forward angles in the center of mass where the direct mechanism is dominant, thus allowing a good description of the angular distribution by the DWBA method.

\begin{figure}[!htpb]
  \includegraphics[clip, trim=0cm 2cm 0cm 3cm, width=\textwidth]{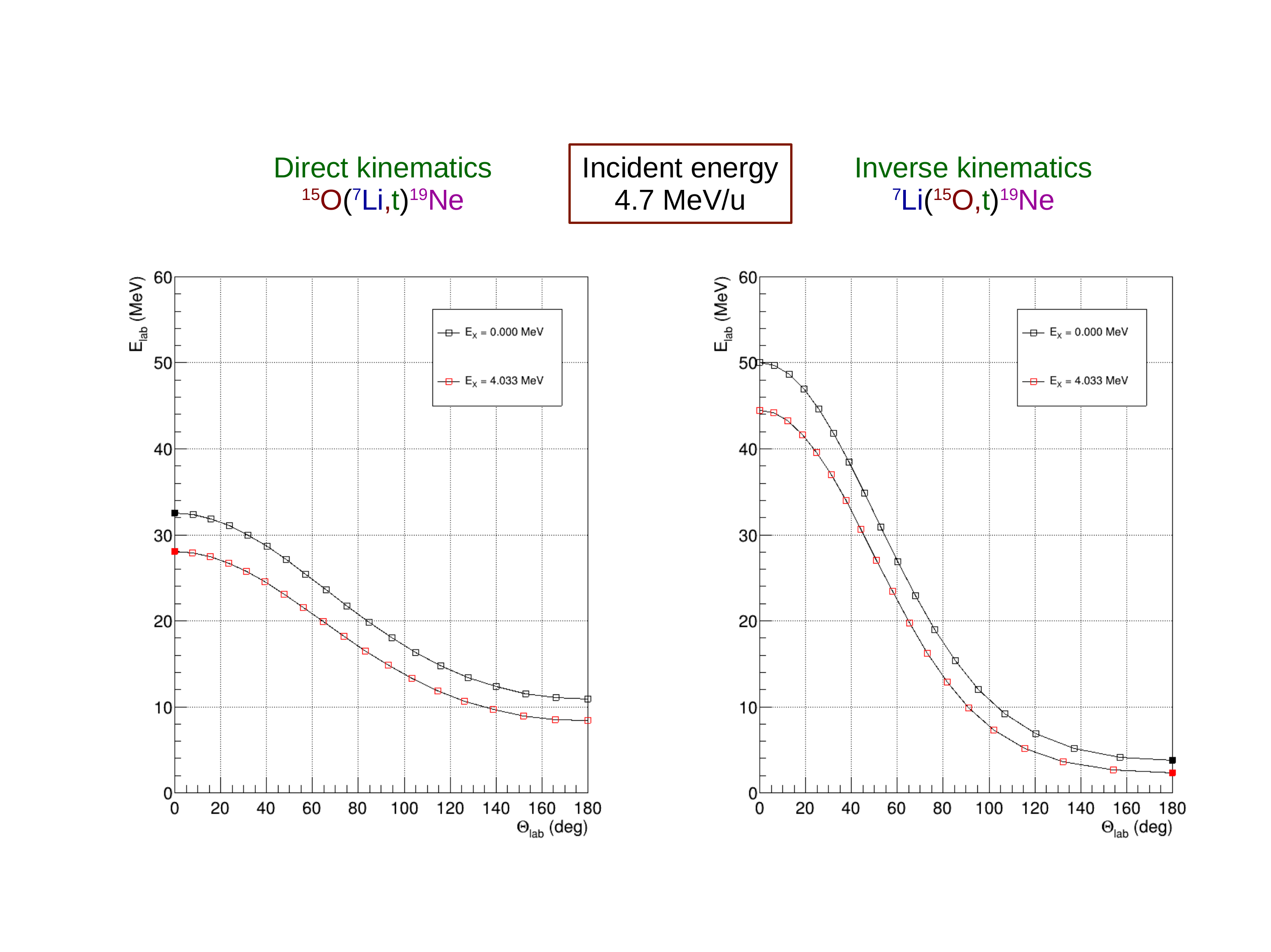}
  \caption{\label{fig:kinematics}
     (Color online) Kinematic calculations for the $^{15}$O($^7$Li,t)$^{19}$Ne reaction showing the triton energy as a function of the laboratory angle in case of direct (left panel) and indirect (right panel) kinematics. Both calculations are performed for the same center of mass energy.}
\end{figure}

While the center of mass frame is best suited for describing the reaction mechanism, the experimental study occurs in the laboratory frame. There are two experimental possibilities to perform a given two-body reaction study: either the projectile is lighter than the target (direct kinematics), or the projectile is heavier than the target (inverse kinematics). Choosing one or the other option will have profound consequences on the nature of the experimental system. To illustrate this point the kinematic lines ($E$ vs $\theta_{lab}$) of the tritons coming out from the $^{15}$O($^7$Li,t)$^{19}$Ne $\alpha$-particle transfer reaction are presented in Fig.~\ref{fig:kinematics} in the case of direct (left panel) and indirect (right panel) kinematics\footnote{Strictly speaking one should not consider the case of direct kinematics for the present reaction since a target containing $^{15}$O nuclei is impossible to produce due to its short half-life of 122.24~s.}. The population of the ground-state and the excited state at 4.033~MeV in $^{15}$O are represented. Square markers are spaced by 10\degree\ in the center of mass and the filled square corresponds to 0\degree.

The first striking difference is that the forward angles in the center of mass correspond to forward/backward angles in the laboratory frame in case of direct/inverse kinematics\footnote{This is usually the case for stripping reactions. In the case of pick-up transfer reactions (e.g. (p,d), (d,$^3$He)...), the forward angles in the center of mass correspond to forward angles in the laboratory frame also in inverse kinematics.}. In addition the tritons have a rather large energy  (about 30~MeV) in case of direct kinematics, while the energy is much smaller (about 3~MeV) in case of inverse kinematics. These two observations will dictate very different experimental setups, and the specifics concerning direct and inverse kinematics studies are now detailed.

\vspace{1 cm}
\subsection{Direct kinematics studies}
Historically transfer reactions were performed using stable beams in direct kinematics. The first detection systems were based on collimated silicon detectors mounted in a $\Delta E-E$ fashion allowing Particle IDentification (PID) based on the energy loss and residual energy deposited in each detector. In such experimental studies several of these telescopes were positioned at different detection angles around the target. Small angles in the laboratory frame were preferred since in direct kinematics forward center of mass angles correspond to forward laboratory angles. One of the limiting aspect of this approach is the intrinsic energy resolution of the silicon detectors which is typically $\approx15-20$~keV. This can be easily superseded by the use of magnetic spectrographs of high resolving power ($E/\Delta E = 3000 - 5000$) such as the Enge Split-Pole~\cite{Spencer1967} or Q3D~\cite{Loffler1973} design (see examples in Sec.~\ref{sec:n13ap} and \ref{sec:si30pg}, respectively). The detection system at the focal plane of the spectrometers usually comprises one (or two) position sensitive detectors recording the magnetic rigidity of the light particles entering the acceptance; a gas detector measuring the energy loss of the particles; and a plastic scintillator where the residual energy of the particle is deposited.

The differential cross section corresponding to a populated state in the residual nucleus is calculated from the light particle yield determined at each detection angle $N_c(\theta_{lab})$ using the following formula:
\begin{equation}
  \left(\frac{d\sigma}{d\Omega}\right)_{c.m.}(\theta_{c.m.}) = 
  \frac{N_c(\theta_{lab})}{Q(\theta_{lab}) N_{target} \Delta\Omega_{lab}} J(\theta_{lab}),
  \label{e:dsig}
\end{equation}
where $Q(\theta_{lab})$ is the accumulated charge at each angle, $N_{target}$ is the number of target atoms per unit area, $\Delta\Omega_{lab}$ is the solid angle, and $J(\theta_{lab})$ is the Jacobian for the laboratory to center-of-mass transformation of the $A(a,c)C$ reaction at each detection angle. 

Transfer reaction studies with stable beams in direct kinematics are rather straightforward. While the spectrometer requires a dedicated hall the complexity of the detection system is usually low with a limited number of electronic channels. The main delicate point in such approach comes from the targets. First because they must be very thin (between tens and hundreds of $\mu$g cm$^{-2}$) in order to limit their contribution to the overall energy resolution budget. They are then extremely delicate to produce and fragile to manipulate. Their purity is another point which deserves a special care because reactions on any other nuclei present in the target may produce unwanted contamination peaks hindering the states of interest. It is then of uttermost importance to have isotopically enriched material when needed, to limit the backing material thickness when the target cannot be self-supported, and to choose carefully the compound form. Concerning the last point and as an example of a transfer reaction on fluorine nuclei, lithium or calcium fluoride targets will not produce the same background, and one or the other compound may be best suited depending on the reaction studied.

\vspace{0.5 cm}
\subsection{Inverse kinematics studies}
The advent of radioactive ion beams (RIBs) allows to perform transfer reaction studies involving nuclei far from the valley of stability (see example in Sec.~\ref{sec:fe60dp}. Beam intensities are much smaller than for stable beams and should be preferably at least 10$^5$~pps in order to perform a transfer reaction study. The beam properties are one of the crucial aspects in such studies, and depending on how the RIB is produced it may be contaminated with other species and have a large emittance. Therefore, detectors tracking the beam position, such as CATS~\cite{Ottini1999}, PPACs~\cite{Kumagai2013} to cite a few of them, are usually used to reconstruct the position of the incident ions at the target location event by event. Identification of the incident beam with respect to other species is also undertaken with standard time of flight and energy loss techniques. In contrast to the direct kinematics case the solid state targets are much more simpler and easier to handle with CH$_2$, CD$_2$ and LiF being mainly used. On the other hand, gas targets can be very complex and usually relies on supersonic gas jet~\cite{Schmidt18} or cryogenics~\cite{Ryuto2005} technology.

The fact that RIB intensities are much smaller than in the case of stable beams experiment has a profound impact on the design of the detection setup. Let us illustrate this point with the case of the $^{15}$O($^7$Li,t)$^{19}$Ne reaction. For a stable beam study of this reaction in direct kinematics we could consider the following typical parameters: a $^7$Li$^{3+}$ beam intensity of about 100 pnA, a target thickness of about 100~$\mu$g.cm$^{-2}$, and a spectrometer solid angle $\Delta\Omega\approx 5$~msr. In an inverse kinematics study of this reaction~\cite{Jenn2019} the $^{15}$O beam intensities are about 10$^7$~pps which is 6 orders of magnitude lower than for the stable beam experiment. The ways of compensating this dramatic loss is to increase the triton detection efficiency and the target thickness. Concerning the charged particle detection system, it is usually based on large acceptance silicon systems where the angular information needed to reconstruct the excitation energy mostly comes from a very high segmentation of the silicon array. Among the many existing arrays, and to only cite a few of them, TIARA~\cite{tiara} uses single-sided silicon stripped detectors, MUST2~\cite{Poll05} and SHARC~\cite{sharc} are based on double-silicon stripped detectors, and ORRUBA~\cite{orruba} relies on resistive silicon detectors. If we consider a close to $2\pi$~sr coverage for these detectors, this is about 3 orders of magnitude more than a typical spectrometer acceptance in direct kinematics. The last parameter than can be tuned to compensate the lower RIB intensities is the target thickness which is typically in the mg.cm$^{-2}$ range, one order of magnitude higher than for direct kinematics studies. The target thickness must be carefully chosen and results from a trade-off between excitation energy resolution and counting rate. In order to mitigate this trade-off it is more and more usual to add in the detection setup a high performance (efficiency and resolution) $\gamma$-ray spectrometer. In that case the reconstructed excitation energy of the residual nucleus based on the charged particle alone does not provide the resolution to separate all populated states, however the coincidence with $\gamma$-rays provides a clean measurement and allows to isolate the contribution of a single state to the excitation energy spectrum. The major silicon arrays cited previously have been coupled to efficient $\gamma$-ray spectrometers such as EXOGAM~\cite{exogam}, AGATA~\cite{agata}, TIGRESS~\cite{tigress} and GRETINA~\cite{gretina}, with the following combinations: TIARA-EXOGAM, MUST2-EXOGAM, MUGAST-AGATA, SHARC-TIGRESS and ORRUBA-GRETINA.

Transfer reaction studies with radioactive ion beams are very challenging and require complex experimental setups. Tremendous progresses over the past 20 years have been made concerning the development of both highly efficient and granular charged particles and $\gamma$-rays spectrometer. Despite of these achievements RIB transfer experiments typically last between one and two weeks with a limited level of accumulated statistics. However this is a unique way to explore regions of the nuclear charts where some of the most extreme astrophysical processes occur.

\section{Examples of experimental transfer reaction studies} \label{sec:experiments}
After some general considerations on the type of transfer reactions useful in nuclear astrophysics, three examples will be presented. The first two examples concern the study of the resonant part of the $^{30}$Si(p,$\gamma$)$^{31}$P and $^{13}$N($\alpha$,p)$^{16}$O reactions studied by means of the one proton ($^3$He,d) reaction, and the $\alpha$-particle ($^7$Li,t) transfer reaction on the mirror reaction, respectively. The last example concerns the study of the direct capture component of the $^{60}$Fe(n,$\gamma$)$^{61}$Fe reaction through the one neutron (d,p) transfer reaction.

\vspace{0.5 cm}
\subsection{Transfer reactions in nuclear astrophysics}
Several transfer reactions can be used to extract spectroscopic factor for the same states of astrophysical interest. The one-proton ($^3$He,d), ($^{4}$He,t) and (d,n) transfer reactions can be used to extract the proton spectroscopic factor of states involved in proton captures reactions. Similarly the one-neutron ($\alpha$,$^{3}$He) and (d,p) transfer reactions can be used to study the resonant and direct components of neutron capture cross-sections. The choice between these transfer reactions is driven by considerations on a good linear and angular momentum matching~\cite{Austern1970}. The transferred linear momentum depends strongly on the beam energy and on the $Q$-value of the transfer reaction. Since in nuclear astrophysics small transferred angular momenta are relevant in most of the cases because of the low associated centrifugal barrier, transfer reactions having a smaller $Q$-value are generally mostly used. As such, the (d,p) and ($^3$He,d) transfer reaction are a very common choice for one-neutron and one-proton transfer reactions, respectively.

In the case of one-proton transfer reactions both the ($^3$He,d) reaction~\cite{Parikh2014,Gillespie2017} and the (d,n) reaction~\cite{Adekola2011} have been used extensively, though the neutron detection may bring some experimental complexity. For the one-neutron transfer case the (d,p) reaction has been mostly used~\cite{deSereville2007, Kozub2012, Giron2017}. Note that the different momentum matching of two reactions transferring the same nucleon can provide useful hints on the nature of the populated states. In that case the same state is populated in a different way according to the reaction, and a distinction between low and high spins may be established (see Ref.~\cite{Wang1989} for a comparison of the ($^3$He,d) and ($^4$He,t) reactions).

The (p,d) and (p,t) pickup reactions are very valuable tools to study proton-rich nuclei of astrophysical interest such as in classical novae and type I X-ray bursts. The Q-values of both reactions are strongly negative, and in case of the (p,t) reaction proton beam energies larger than 30 MeV are often needed favoring the use of cyclotron instead of electrostatic accelerators. The (p,d) direct reaction mechanism can be well described by the DWBA formalism and it is then possible to extract useful spectroscopic information from the analysis of the angular distributions~\cite{Bardayan2015}. This is more complicated in the case of (p,t) reactions since the two neutrons can be transferred as a pair in a single step or in the possible two steps (p,d)(d,t) path which requires to know the spectroscopic factor and energy of the intermediate states. This makes the analysis of the angular distribution more delicate~\cite{Bardayan2006} and not as reliable as a single particle transfer reaction. Despite these complications the (p,t) reaction is widely used because of its selectivity which mainly populates natural spin and parity states (if a single step is assumed) of even-even nuclei. 

Alpha-particle transfer reactions are very useful to study the spectroscopy of nuclei involved in $\alpha$-induced reactions such as ($\alpha$,$\gamma$), ($\alpha$,n) and ($\alpha$,p) reactions in helium rich environments. The most generally used transfer reactions are the ($^6$Li,d) and ($^7$Li,t) reactions. At the time of early studies the ($^{6}$Li,d) reaction was used extensively because the $L=0$ relative motion of the $\alpha$-particle and deuteron in $^{6}$Li allowed for a Zero-Range DWBA treatment. This approximation is not correct in the case of the ($^7$Li,t) reaction for which the $\alpha$-particle and triton are in a $L=1$ relative motion. Multi-step effects may be important in $\alpha$-particle transfer reactions \cite{Keel03}, however a comparison of the two transfer reactions off $^{12}$C has shown that these effects are reduced when using a $^{7}$Li beam~\cite{Bech78,Kemp80}. In addition the transfer cross-sections to low spin states are enhanced in case of the ($^7$Li,t) transfer reaction due probably to the non-zero $\alpha$-particle angular momentum in $^7$Li~\cite{Bethg70}, and the angular distributions show much stronger direct features by exhibiting more forward pronounced maxima~\cite{Bethge70}. However the angular distributions have less oscillatory structures than those from ($^6$Li,d) due to the fact that the $\alpha$+t cluster in $^7$Li exists in a relative $p$-state which implies two transferred $\ell$ values that superimpose to form a state of a given spin and parity in the final nucleus~\cite{Bethge70}. According to~\cite{Bethge70}, in comparison to ($^6$Li,d), ($^7$Li,t) transfer reaction seems to populate more selectively states with $\alpha$ structure.

\vspace{1 cm}
\subsection{Case of the $^{30}$Si(p,$\gamma$)$^{31}$P reaction} 
\label{sec:si30pg}
Globular clusters are vital testing grounds for models of stellar evolution and the early stages of the formation of galaxies. Abundance anomalies such as the enhancement of potassium and depletion of magnesium have been reported in the globular cluster NGC 2419~\cite{Cohen2012}. They can be explained in terms of an earlier generation of stars polluting the presently observed stars, however, the nature and properties of the polluting sites is not clear (see Ref.~\cite{Charbonnel2016,Bastian2018} for a review). It has been shown that the potential range of temperatures and densities of the polluting sites depends on the strength of a number of critical reaction rates including $^{30}$Si($p,\gamma$)$^{31}$P~\cite{Dermigny2017}.

Several resonances are known in the Gamow window (\Ercm{100-500} range) associated to the temperature range of interest between 100~MK and 250~MK. Their fractional contribution to the $^{30}$Si($p,\gamma$)$^{31}$P reaction rate have been calculated and several resonances have been identified as dominating the reaction rate~\cite{Dermigny2017}. While the strength of the highest resonances having $E_r^{c.m.}>400$~keV can be accessed by direct measurement this is not the case for lower energy resonances because of the much smaller barrier penetrability. Since the partial proton width of the corresponding $^{31}$P state is expected to be much smaller than the radiative partial width, the resonance strength is then proportional to the particle width. As described in Section~\ref{sec:res} the proton partial width is related to the proton spectroscopic factor which can be in turn determined from a one-proton transfer reaction. The shape of the angular distribution will give some precious insights on the angular momentum of the transferred proton $\ell$ which will allow to constrain the unknown spin-parity of the resonances to $J=\ell\pm1/2$.

\begin{figure*}[!htpb]
  \includegraphics[width=1.02\textwidth]{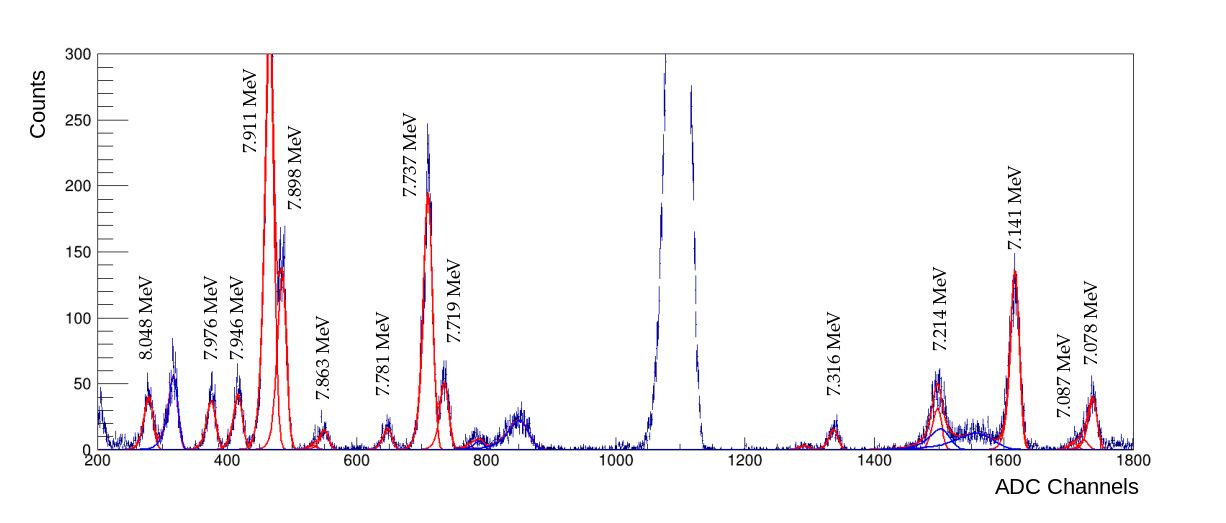}
  \caption{\label{fig:Q3D}
(Color online) Deuteron magnetic rigidity spectrum at a spectrometer angle of 16\degree. Excitation energies in $^{31}$P between 7.0 and 8.1~MeV are covered. The best fit of the spectrum is shown together with individual contributions for $^{31}$P states (red) and contamination peaks (blue).}
\end{figure*}

The importance of low-lying resonances above the p+$^{30}$Si threshold ($S_p=7297$~keV) has been investigated using the one-proton $^{30}$Si($^3$He,d)$^{31}$P transfer reaction. The experiment was performed with a $^3$He$^{2+}$ beam of about 200 enA accelerated to 25~MeV by the TANDEM accelerator of the Maier-Leibniz-Laboratory at Munich. The beam impinged a thin target (20~$\mu$g/cm$^2$) of enriched silicon oxyde located at the object focal point of the Q3D magnetic spectrometer~\cite{Loffler1973}. The deuterons were momentum analysed and detected by the focal plane system allowing their clear identification from other light particles, and the measurement of their magnetic rigidity. A typical deuteron magnetic rigidity spectrum at a spectrometer angle $\theta=16$\degree\ is displayed in Figure~\ref{fig:Q3D}. Since the Q3D magnetic spectrometer has been tuned to cancel the kinematic broadening of the $^{30}$Si($^3$He,d)$^{31}$P reaction, narrow peaks are associated to $^{31}$P states (red components) while the broad structures (unlabeled blue components) correspond to reactions on other target elements such as $^{12}$C and $^{16}$O. The extremely good energy resolution of about 6--7~keV (FWHM) in the center-of-mass allows a clear separation of the two components of the 7719- and 7737-keV doublet relevant in the present study.

\begin{figure*}[!htpb]
  \includegraphics[width=1.02\textwidth]{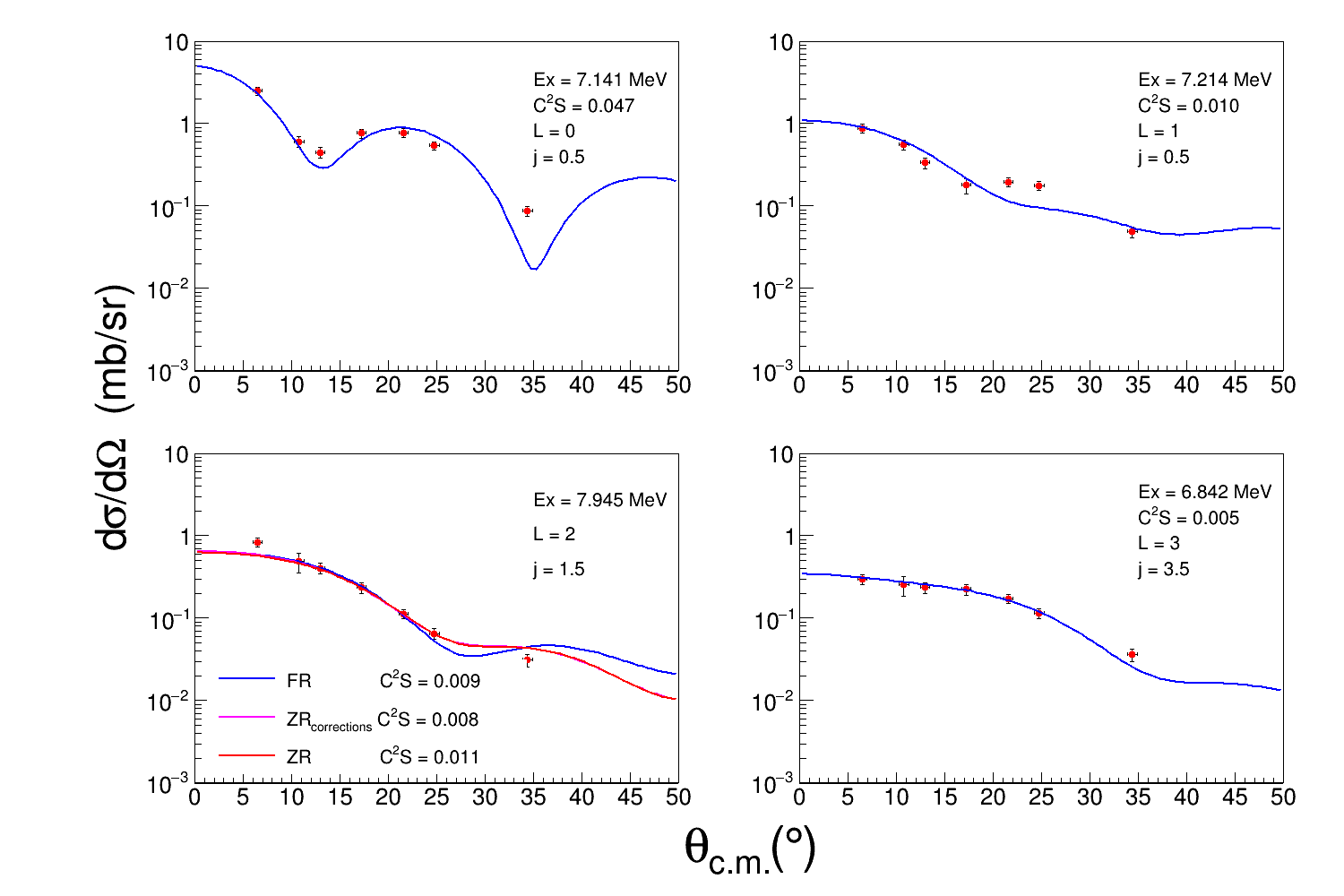}
  \caption{\label{fig:dsig31P}
     (Color online) Selection of experimental differential cross-sections of $^{31}$P states populated with the $^{30}$Si($^3$He,d)$^{31}$P transfer reaction. Each panel correspond to a different transferred relative angular momentum, and the blue solid lines represent finite-range DWBA calculations normalized to the data. Zero-range calculations are also reported in case of the 7.945~MeV state (see text for details).}
\end{figure*}

The differential cross-sections corresponding to populated $^{31}$P states were calculated from the deuteron yield determined at each spectrometer angle, and examples are shown in Figure~\ref{fig:dsig31P} for states populated by different transferred angular momentum~\cite{Harrouz2020}. In all cases the rapidly varying cross-section on a limited forward center-of-mass angular range is indicative of states which are populated through a direct mechanism. The differential cross-sections also have a very characteristic shape which depends on the transferred angular momentum, e.g. the position of the first minimum of the cross-section increases with the magnitude of the transferred angular momentum. Finite-range DWBA calculations performed with the FRESCO code~\cite{Thom88} are represented in blue for a selection of bound and unbound states, and a very good agreement is obtained with the experimental data. For unbound states the weakly bound approximation is assumed, and a bound form factor corresponding to a state bound by 10~keV is considered.

While the optical potential parameters for the entrance channel come from an experimental study of the same transfer reaction at the same bombarding energy~\cite{Vernotte1990}, the parameters for the exit channel come from set F of Daehnick~et~al. global deuteron potentials~\cite{Dae80}. The proton form factor was obtained by adjusting the depth of a standard Woods-Saxon well in order to reproduce the experimental proton separation energy of each $^{31}$P state. The geometry of the well had a radius and diffuseness of $r_0=1.25$~fm and $a_0=0.65$~fm, respectively. 

The finite-range calculations have been performed using the $\langle^3$He$\mid$d$\rangle$ overlap obtained by the Green's function Monte Carlo method using the Argonne $\nu_{18}$ two-nucleon and Illinois-7 three-nucleon interactions~\cite{Brida2011}. The shape of the angular distribution is very similar to the case of a zero-range DWBA calculation (red curve) and the forward angles are similarly well described. A difference of 20$\%$ is observed between the spectroscopic factors extracted with both calculations. A zero-range DWBA calculation has also been performed using non-local and finite-range corrections (see Ref.~\cite{Vernotte1990,Bassel1966} for the parameters) for comparison, and a spectroscopic factor lower by 30\% with respect to a zero-range calculation is obtained in agreement with previous work~\cite{Vernotte1990,Bassel1966}. Note that with modern computational resources finite-range DWBA calculations should always be preferred.

The proton spectroscopic factors obtained from the finite-range DWBA analysis have been used to derive the proton partial widths using Equation~\ref{eq:partialWidth}. The weakly-bound approximation was assumed and the validity of this assumption was explored by performing zero-range DWBA calculations to unbound states with the DWUCK4 code~\cite{Kunz} which relies on the Vincent and Fortune complex integration procedure of the radial integrals~\cite{Vincent1970}. In addition, DWUCK4 calculates the proton partial widths and a maximum difference of 15\% was observed with the weakly-bound approximation. This is related to the fact that the proton wave functions, even for unbound levels, are well described in the weakly-bound approximation because the high Coulomb barrier leads to a strong suppression of the wave function at large radii.

\vspace{0.5 cm}
\subsection{Case of the $^{13}$N($\alpha$,p)$^{16}$O reaction} \label{sec:n13ap}
It has been recently suggested that hydrogen ingestion into the helium shell of massive stars could lead to high $^{13}$C and $^{15}$N excesses when the shock of a core-collapse supernova (CCSN) passes through its helium shell~\cite{Pignatari2015}. This prediction questions the origin of extremely high $^{13}$C and $^{15}$N abundances observed in rare presolar SiC grains which is usually attributed to classical novae~\cite{Amari2001}. In this context the \nap\ reaction plays an important role since it is in competition with $^{13}$N $\beta^+$-decay to $^{13}$C.

The evaluation of the \nap\ reaction rate in the temperature range of interest between 0.4 and 1~GK requires a detailed knowledge of the structure of the compound nucleus $^{17}$F within around 2.5~MeV above the $^{13}$N+$\alpha$ threshold. Spins and parities are known in most cases and the energy and total widths of the states are known experimentally~\cite{Til93}. Given that the $^{13}$N+$\alpha$ threshold ($S_{\alpha}$~=~5818.7~(4)~keV) is much higher than the $^{16}$O+p threshold ($S_p$~=~600.27~(25)~keV), the states in the region of interest decay mainly by proton emission, so that $\Gamma_p \approx \Gamma_{tot}$. Their contribution to the reaction rate is therefore directly proportional to their unknown alpha-particle widths, which can be calculated from the spectroscopic factors obtained in an adequate $\alpha$-particle transfer reaction.

The $^{13}$N($^7$Li,t)$^{17}$F transfer reaction would be the most evident reaction to perform. However, while not impossible, such an experimental study in inverse kinematics would require the use of an intense radioactive $^{13}$N beam ($>10^7$~pps) with a complex setup including an identification station at 0\degree, a large coverage charged particle array for the triton detection and an efficient $\gamma$-ray array needed to cope with the relatively high $^{17}$F level density. Such setups have recently been used for the study of the $^7$Li($^{15}$O,t$\gamma$)$^{19}$Ne reaction at GANIL~\cite{Jenn2019} and the $^7$Li($^{17}$O,t$\gamma$)$^{21}$Ne reaction at TRIUMF~\cite{Williams2019}. Given that, the $\alpha$-particle widths of the $^{17}$F states were deduced from the properties of $^{17}$O analog states when such a correspondence is established~\cite{Meyer2020}. The \clt\ reaction measurement~\cite{Pel08} using a stable beam in direct kinematics was performed at the Tandem-ALTO facility in Orsay, France. A 34-MeV $^7$Li$^{3+}$ beam of about 100~enA impinged an enriched $^{13}$C target of 80~$\mu$g/cm$^2$, and the tritons were momentum analysed and focused on the focal-plane detection system of an Enge Split-Pole spectrometer~\cite{Spencer1967}. The energy resolution of about 50~keV (FWHM) allowed to separate all the states of interest (see Figure~2 in Ref.~\cite{Meyer2020}) and to extract their angular distributions.

\begin{figure*}[!htpb]
  \includegraphics[width=1.02\textwidth]{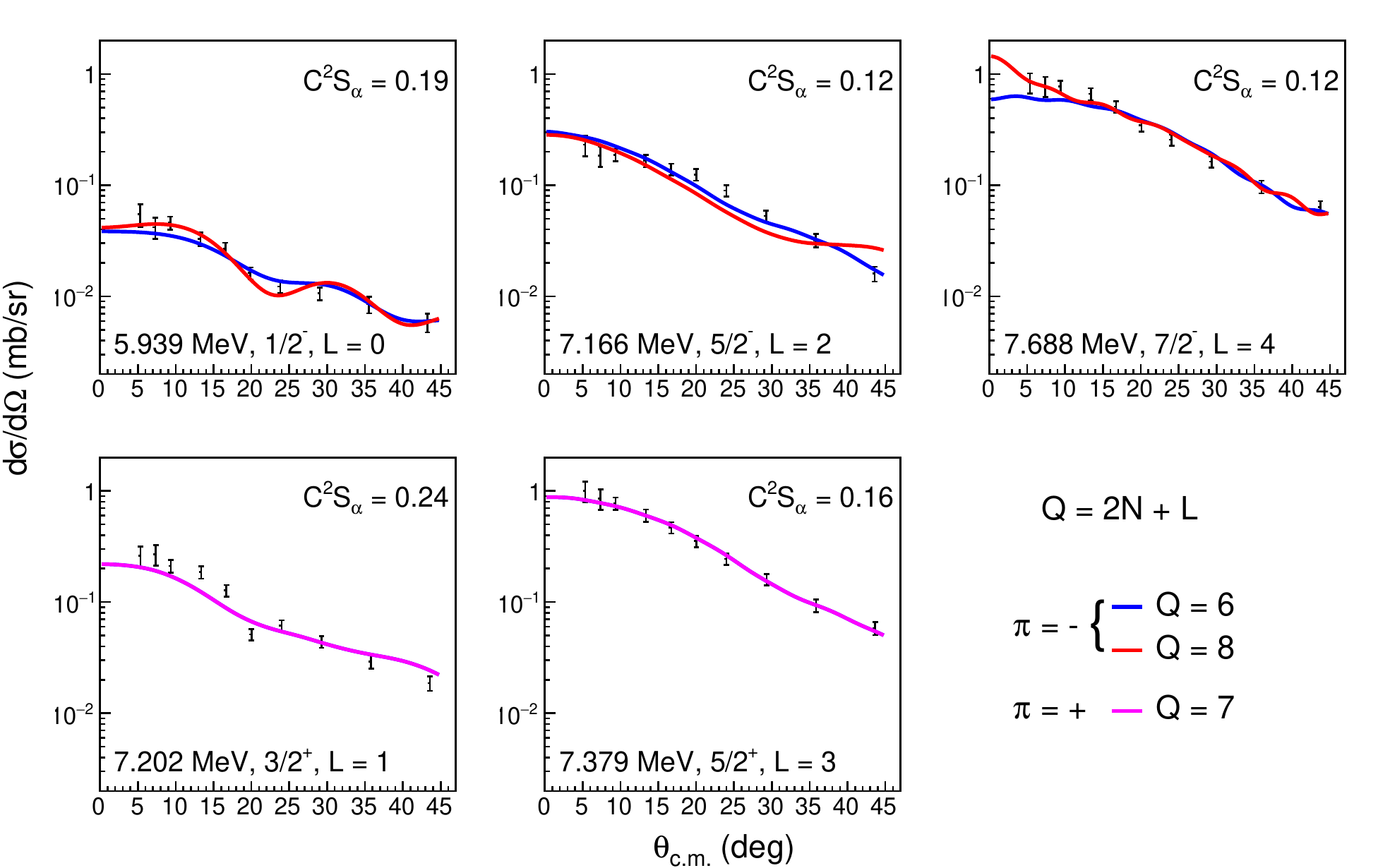}
  \caption{\label{fig:dsig17O}
     (Color online) Selection of experimental differential cross-sections of $^{17}$O states populated with the \clt\ transfer reaction. Negative- and positive-parity states are in the upper and lower row, respectively. Solid lines represent finite-range DWBA calculations normalized to the data for different values for the number of quanta $Q$ in the relative $\alpha$+$^{13}$C motion (see text for more details).}
\end{figure*}

Examples of differential cross-sections for positive and negative parity $^{17}$O states populated with different transferred angular momentum $L$  are shown in Figure~\ref{fig:dsig17O}, together with finite-range DWBA calculations performed with the FRESCO code. An excellent agreement is observed between the theory and the experiment which supports a single step direct mechanism for the population of $^{17}$O states using the \clt\ reaction. However unlike the single nucleon transfer reactions the angular distributions obtained from ($^7$Li,t) reactions are usually less pronounced with much less marked angular minima and maxima. Also, the shape of the angular distributions is not so sensitive to the transferred angular momentum $L$ as can be observed in Figure~\ref{fig:dsig17O} which makes its determination more delicate.

Details about the ingredients needed for the FR-DWBA calculations, such as the optical potential parameters, the overlap between the $\alpha$+t and $^7$Li systems, and the geometry of the Woods-Saxon potential used to compute the wave-function describing the $\alpha$+$^{13}$C relative motion, can be found in Ref.~\cite{Meyer2020}. An important ingredient is the number of nodes $N$ (defined here as excluding the origin) of the radial part of the $\alpha$+$^{13}$C wave-function. Even though there is a limited sensitivity of the angular distributions to the number of nodes $N$, its determination should be whenever possible guided by microscopic considerations such as a cluster description of the states or from the insight of shell-model calculations. The link between these two views is not straightforward but there have been recent efforts to perform FR-DWBA calculations of the $^{12}$C($^7$Li,t)$^{16}$O reaction with the $(N,L)$ parameters for the various components of the $\alpha$+$^{12}$C wave functions obtained from shell model calculations~\cite{Weintraub2019}. In case of $^{17}$O while lower states are well described by a single neutron above a close $^{16}$O core, high energy states have been theoretically predicted~\cite{Brown1966}, and observed for some of them~\cite{Cunsolo1981}, to be dominated by a mixture of 2p-1h/4p-3h and 3p-2h/5p-4h configurations\footnote{where the $x$p-$y$h notation has the usual meaning of $x$ particles in the $sd$-shell and $y$ holes in the $1p$ orbitals.} for negative- and positive-parity $^{17}$O states, respectively. In the following discussion we make use of the number of quanta $Q=2N+L$ for the relative $\alpha$+$^{13}$C motion.

In the case of positive-parity states in $^{17}$O, the 5p-4h configuration is not expected to be populated in direct $\alpha$-particle transfer since the shell model overlap between a 5p-4h configuration in $^{17}$O and the $^{13}$C in its ground state and an $\alpha$-particle would be zero. Since the experimental angular distributions are well described by a one step direct reaction mechanism this would mean that the reaction mechanism is sensitive to the 3p-2h configuration, either because the states have a dominant 3p-2h configuration, or, if they have a dominant 5p-4h configuration, because the ($^7$Li,t) mechanism is sensitive to the small admixture of 3p-2h configuration. When the two neutrons and protons of the transferred $\alpha$-particle are positioned on the orbitals respecting a 3p-2h configuration one obtains $Q=7$ for the number of quanta in the relative $\alpha$+$^{13}$C motion from which can be derived the number of nodes $N$ of the radial part of the wave function using the Talmi-Moshinsky relation~\cite{Mos59} (see Section~\ref{sec:ff}).

In a similar way for negative-parity states in $^{17}$O, the 2p-1h and 4p-3h configurations can be associated to the number of quanta $Q=6$ and $Q=8$, respectively. Both cases are displayed in Figure~\ref{fig:dsig17O} and the two calculations show very similar behaviours except for angles larger than 35\degree. This emphasizes that the shape of the angular distributions in DWBA are not particularly sensitive to the number of nodes $N$. In the present case, without a better knowledge of the $^{17}$O states structure, a choice has therefore to be made for $Q$ and hence $N$. The value of the number of nodes has a strong impact on the alpha spectroscopic factors which can vary by a factor of almost two as reported in Tab.~\ref{tab:Q17O} for a few states. However this has a much limited impact on the determination of the alpha width which only varies by less than ten percent in the present case. This is explained by the fact that when computing the reduced $\alpha$-particle width (see Equation.~\ref{eq:reducedWidth}), there is a compensation effect between the spectroscopic factor and the radial part of the $\alpha$+$^{13}$C wave function whose shape depends strongly on its number of nodes. Similar effects were also observed in the study of the $^{15}$N($^7$Li,t)$^{19}$F reaction~\cite{deOliveira1996}.

\begin{table*}[!htpb]
  \centering
  \caption{\label{tab:Q17O}
     Alpha-particle spectroscopic factors and widths for negative-parity $^{17}$O states obtained when considering a number of quanta in the relative $\alpha$+$^{13}$C motion $Q=6$ and $Q=8$. Comparison with alpha widths from the literature is provided.}
  \begin{tabular}{@{\extracolsep{4pt}}ccccccccc@{}}
    \hline \hline
    \multicolumn{3}{c}{NNDC~\cite{Til93}} & \multicolumn{3}{c}{$Q=6$} & \multicolumn{3}{c}{$Q=8$} \\ \cline{1-3} \cline{4-6} \cline{7-9}
    $E_x$ & $J^\pi$ & $\Gamma_\alpha$ & $N, L$ & $C^2S_\alpha$ & $\Gamma_\alpha$ & $N, L$ & $C^2S_\alpha$ & $\Gamma_\alpha$ \\
    (keV) &         &  (keV)          &       & & (keV)        &       &  & (keV) \\ \hline
    5939 (4) & 1/2$^-$ &        & 3, 0 & 0.19 &                      & 4, 0 & 0.12 & \\
    7166 (8) & 5/2$^-$ & 3.3$\times$10$^{-3}$ & 2, 2 & 0.12  & 3.4$\times$10$^{-3}$ & 3, 2 & 0.074 & 3.7$\times$10$^{-3}$ \\
    7688 (9) & 7/2$^-$ & 1.0$\times$10$^{-2}$   & 1, 4 & 0.12  & 3.3$\times$10$^{-3}$ & 2, 4 & 0.055 & 3.5$\times$10$^{-3}$ \\
    \hline \hline
  \end{tabular}                                                                
  \footnotetext[1] {The quantities $N$ and $L$ are the radial nodes (including the origin) and orbital 
  angular momentum assigned to the center of mass motion of the 
  $\alpha$-cluster in $^{17}$O.}
  \footnotetext[2] {$\Gamma_\alpha = 2P_l(a,E)\frac{\hbar^2a}{2\mu} 
  C^2S_\alpha\; |\phi(a)|^2$ with $|\phi(a)|$ being the radial part of the 
  $^{13}$C+$\alpha$ wave function evaluated at the channel radius $a =7.5$~fm (see text).}
 \footnotetext[3] {From~\cite{Pel08}, the reduced width $\gamma^2_\alpha$ is given instead of $\Gamma_\alpha$.}
  \footnotetext[4] {This doublet is not resolved experimentally so the deduced 
  spectroscopic factor assumes all the strength is on one or the other state.}
\end{table*}

The $^{13}$N($\alpha$,p)$^{16}$O reaction rate was calculated based on the previous spectroscopic information, and it was found to be within a factor of two of the previous evaluation done by Caughlan and Fowler~\cite{CF88}. A detailed Monte-Carlo study was then used to propagate the nuclear uncertainties to the reaction rate, and a factor of uncertainty of two to three was obtained. This translates into an overall uncertainty in the $^{13}$C production of a factor of 50 when using the lower and upper reaction rates~\cite{Meyer2020}.

\vspace*{0.5 cm}
\subsection{Case of the $^{60}$Fe(n,$\gamma$)$^{61}$Fe reaction} \label{sec:fe60dp} 
$^{60}$Fe(n,$\gamma$)$^{61}$Fe plays an important role in the abundance of $^{60}$Fe which characteristic gamma-ray lines at 1173.23 and 1332.44 keV coming from the decay-chain of $^{60}$Fe-$^{60}$Co-$^{60}$Ni have been observed by the spacecrafts missions RHESSI in 2004 \cite{Smith04} and INTEGRAL in 2007 \cite{Wang07}. The observation of these gamma-ray lines indicates that the nucleosynthesis of $^{60}$Fe is still active in Galaxy since its lifetime 2.6 million years is much smaller than the galactic time evolution which is around 10 million years. An excess of $^{60}$Fe has also been observed in deep ocean crusts and sediments as well as in lunar soils\cite{Knie04, Wall16,Fimi16} and in galactic cosmic rays (CRIS/ACE)~\cite{Binns16}. All these observations have underlined the need for accurate nuclear information concerning the stellar nucleosynthesis and destruction of this nucleus. $^{60}$Fe is mainly produced in massive stars through the weak $s$-process component and it is released in the interstellar medium by the subsequent core-collapse supernovae explosion \cite{Limo06}. Thus, all $^{60}$Fe observations give the opportunity to test stellar models that describe the evolution of massive stars. However, the important uncertainties surrounding the cross-section of the destruction reaction $^{60}$Fe(n,$\gamma$)$^{61}$Fe imply large uncertainties on the predictions of $^{60}$Fe abundance by stellar models.

The direct measurement of the cross-section of this reaction is very challenging due to the radioactive nature of $^{60}$Fe. An alternative method would be to determine the cross-section through the activation method, which was performed by Uberseder et al.~\cite{Ubers09} or by the (d,p) transfer reaction to determine the excitation energies, orbital angular momenta and neutron spectroscopic factors of $^{61}$Fe states that are important for the calculation of the direct component (Section~\ref{sec:direct}) of the (n,$\gamma$) reaction cross-section in the region of astrophysical interest (E$_{c.m}\simeq$30 keV). This method was chosen by Giron et al. \cite{Giron17} to study the strength of the contribution of the direct component to the $^{60}$Fe(n,$\gamma$)$^{61}$Fe reaction. 

Since $^{60}$Fe is radioactive, it is very difficult to produce an $^{60}$Fe target with enough areal density to perform the (d,p) reaction measurement with a deuteron beam.  Consequently, the $^{60}$Fe(d,p)$^{61}$Fe  measurement was performed in inverse kinematics \cite{Giron17}, using the 27 A.MeV $^{60}$Fe secondary beam produced by fragmentation at LISE spectrometer line of GANIL and a deuterated polypropylene CD$_2$ target of 2.6~mg/cm$^2$ to induce the reaction.

The $^{60}$Fe beam intensity produced was of about 10$^5$ pps which is the usual beam intensities one can get with radioactive beams not far from the valley of stability. As discussed in Sec.~\ref{sec:expneeds} these low intensities required the use of large area and highly segmented silicon strip  detector arrays placed at backward angles in the laboratory; four MUST2 telescopes~\cite{Poll05} and an S1 annular DSSSD from Micron Semiconductor Ltd., in order to increase the angular coverage of the protons detection (from 2$^\circ$ to 23$^\circ$ in center of mass) and hence the statistics. 
The $^{60}$Fe beam being produced by fragmentation has a large emittance. Therefore, to determine precisely the location of the proton emission point on the target and its emission angle, two multi-wire proportional chambers (MWPC) called CATS were used to track the beam. To disentangle the different populated states in $^{61}$Fe that can not be discriminated with particle detection, 4 Germanium clovers (EXOGAM)~\cite{exogam} were used to detect the emitted  $\gamma$-rays from the decay of the populated states in $^{61}$Fe. As for the residual fragments, they were identified in mass and charge using their energy loss in the ionization chamber and the time of flight between the plastic scintillator at the end of the line and one of the CATS detectors.

\begin{figure}[h]
\begin{center}
\includegraphics[width=6 cm]{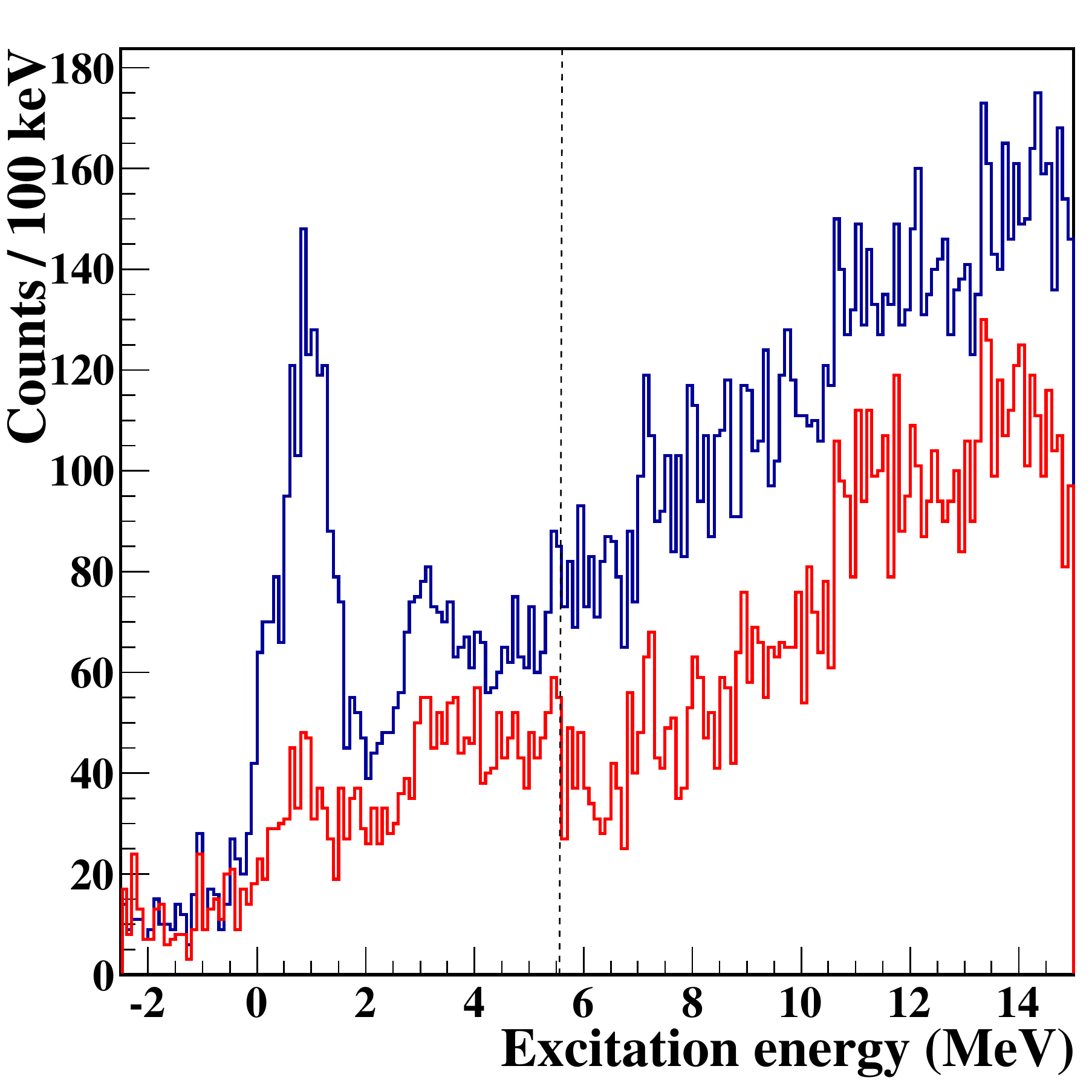}
\includegraphics[width=6 cm]{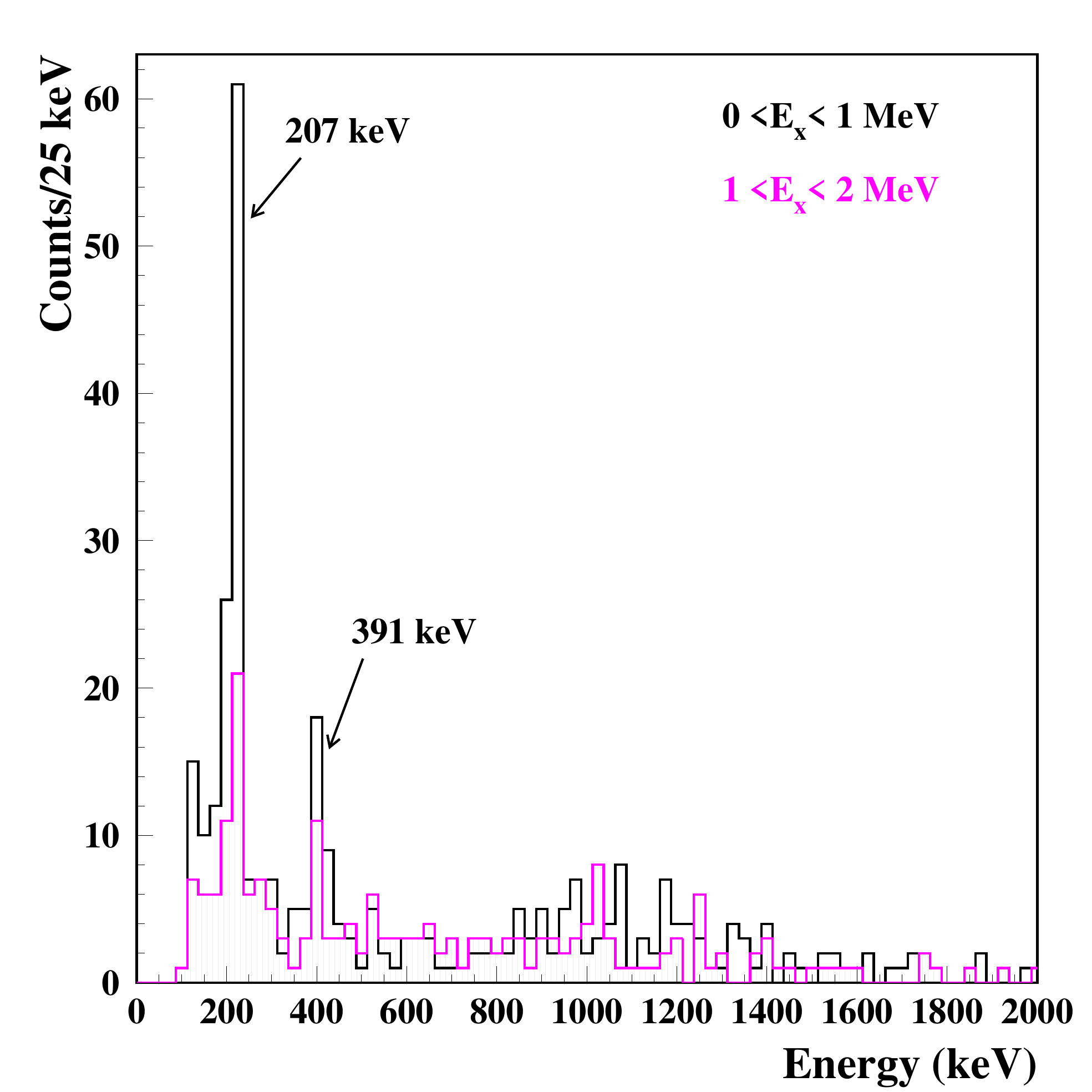}
\caption{(Color online) Left: Measured $^{61}$Fe excitation energy spectrum in coincidences with gammas (red curve) and without 
coincidences (blue curve). The vertical dashed line corresponds to the neutron threshold. Right: Energy spectrum of the $\gamma$-rays in coincidences with protons detected in MUST2 or S1 detectors: Black, for an excitation energy gate 
between 0 to 1 MeV ; pink, for Ex gate between 1 to 2 MeV}
\label{f:Ex_61Feg}
\end{center}
\end{figure}

The reconstructed $^{61}$Fe energy spectrum using MUST2 energy and angle measurements is displayed in Figure~\ref{f:Ex_61Feg} (left panel), 
with and without $\gamma$ coincidences.  

Two peaks are observed below the neutron threshold S$_n$ = 5.58 MeV.
The first peak is around 1 MeV and the other at 3 MeV. The width of these peaks is 1.5 MeV and 2 MeV respectively which 
is much larger than the expected excitation energy resolution, namely 800 keV. This is an indication that several levels are present in the peaks observed. The importance of detecting the $\gamma$-rays is obvious in this case.

One can also observe a drop of about a factor three in the counts of the main peak around 1 MeV when comparing the excitation energy spectrum with and without $\gamma$-ray coincidence while it is between 1.6 to a factor 2 everywhere else. This is a strong indication of the 
population of the isomeric state at 861 keV whose $\gamma$-rays can
not be detected because they are emitted when $^{61}$Fe ions are stopped in the plastic which is at a far distance from the EXOGAM detectors. Indeed the lifetime of the isomeric state ($\tau$=239 ns) is much longer than the time of flight of $^{61}$Fe ions ($\simeq$ 13 ns) between the CD$_2$ target and the plastic.

From the observation of the gamma-ray spectra corresponding to two energy gates in the first peak, from 0 to 1 MeV and 
from 1 to 2 MeV in Figure~\ref{f:Ex_61Feg} (right panel) and from the comparison of the excitation energy spectrum with and without $\gamma$-ray coincidence in Figure~\ref{f:Ex_61Feg} (left panel), three states were clearly identified: the known 207 keV, the 391 keV and the 
isomeric state at 861 keV. 

To extract the proton angular distributions of the identified states, a deconvolution of the first peak observed in $^{61}$Fe excitation energy spectrum around 1 MeV was performed considering the ground state (gs), the three well identified populated states at 207 keV (J$^\pi$=5/2$^-$), 391 (J$^\pi$=1/2$^-$) and 
861 keV (J$^\pi$=9/2$^+$) and also a higher level centered at 1600 keV representing a mixture of the non-identified higher 
states between 1.2 MeV and 2 MeV \cite{Giron17}. 

\begin{figure}[h]
\begin{center}
\includegraphics[width=9cm]{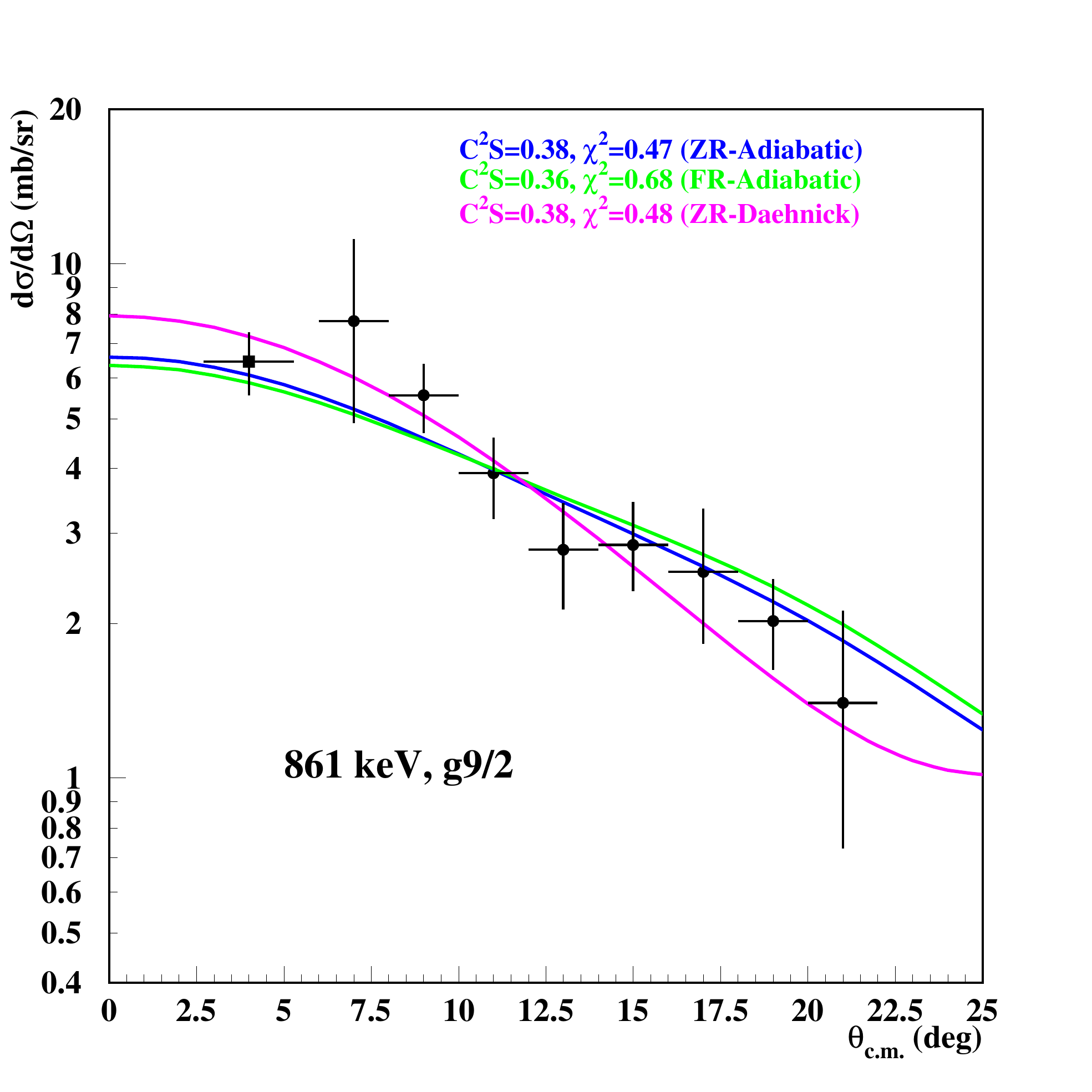}
\caption{(Color online) Experimental differential cross-sections for the 861 keV excited state,  together 
with the different calculations normalized to the data. See text for details.}
\label{f:dsigma_61Fe}
\end{center}
\end{figure}

An example of the extracted proton angular distributions is displayed in Figure~\ref{f:dsigma_61Fe} for the 861 keV state. The blue and the green curves are zero-range (ZR) and finite-range (FR) Adiabatic Distorded Wave Approximation (ADWA) calculations, respectively (see Section 3.1.4). The magenta curve does not take into account the deuteron breakup, and is a zero range calculation using Daehnick et al. global parametrization~\cite{Dae80} for the optical potential describing the entrance channel. All calculations were performed with the FRESCO code. Given the large statistical uncertainties all three calculations give a similar reduced chi-square. However, the incident beam energy of 27 A.MeV
would correspond to an incident deuteron energy of 54 MeV, and deuteron breakup should then be considered, therefore favoring the ADWA calculations. The main effect of taking into account the deuteron breakup is a noticeable difference in the shape of the differential cross-section with respect to the DWBA calculation, e.g. different position of the first angular minimum. Note as well that, in the present case, the ZR- and FR-ADWA calculations give similar results both in terms of the shape of the differential cross-section, and of the spectroscopic factors which differs by only 5\%.

A comparison between the C$^2$S obtained in this work \cite{Giron17} with those predicted by shell-model calculations within a $fpgd$ valence space using the LNPS\footnote{Lenzi, Nowacki, Poves and Sieja} effective interaction \cite{Lenz10} shows a very good agreement within the experimental error bars between the experimental results and the calculations (see  Tab.~\ref{tab:60Feshell}). This confirms further the reliability of the LNPS shell-model calculations in the mass region around N=34.

\begin{table*}[!htpb]
  \centering
  \caption{\label{tab:60Feshell}
 Comparison of the spectroscopic factors obtained in this work and those prediced in LNPS shell-model calculations}
 \begin{tabular}{ c  c  c }
 \hline 
 States & C$^2$S &  C$^2$S \\
   keV  & This work & Shell-model\\
 \hline
 g.s & 0.15$\pm$0.06 & 0.07 \\
 270 & 0.34$\pm$0.10 & 0.42 \\
 391 & 0.58$\pm$0.20 & 0.71 \\
 861 & 0.38$\pm$0.07 & 0.52 \\
 \hline
  \end{tabular}                                                                
\end{table*}

The direct component of $^{60}$Fe(n,$\gamma$)$^{61}$Fe was calculated using the experimental C$^2$S  for the first four excited states of $^{61}$Fe and its value was found to be 0.2 mb at 25 keV. This represents 2$\%$ of the total cross-section measured in \cite{Ubers09}. 

\section{Summary and perspectives}
In this review, we have focused in the transfer reaction method which has been widely used to derive very useful spectroscopic information (spectroscopic factors, partial widths, orbital momenta and resonance energies) needed to evaluate resonant and non-resonant reaction rates of astrophysical interest. 
The theoretical description of the method has been recalled and a review of its use in some recent experimental studies using stable and radioactive beams with different detection systems has been given. 

The current development of exotic radioactive ion beams in many facilities around the world opens new opportunities for the study of astrophysical processes involving nuclei far from the valley of stability, such as the $r$- and $rp$-processes for examples. While transfer reactions in inverse kinematics have been performed with radioactive species since many years, their limited production rates always pushed forward the development of efficient detection systems. The implementation of such systems now relies on coupling in a compact way state of the art charged particles and $\gamma$-ray arrays; the design being always driven by a compromise between $\gamma$-ray detection efficiency and excitation energy resolution obtained from the charged-particles array. Recent examples are the MUGAST silicon array~\cite{mugast} coupled to the AGATA $\gamma$-ray spectrometer~\cite{agata} at GANIL, and annular silicon detector coupled with the TIGRESS $\gamma$-ray array~\cite{tigress} at TRIUMF, to cite a few of them. Owing to the inverse kinematics of such measurements, the fusion and evaporation between the beam and the target usually induces a large background which must be coped with. Several experimental efforts have been focused on the development of dedicated targets limiting such induced background and energy straggling such as the JENSA windowless supersonic jet gas target~\cite{Schmidt18}.

On the theoretical side, many progresses have been made to describe one-nucleon overlap functions as well as to understand the three-body dynamics related to the deuteron breakup degrees of freedom, including the nonlocality effects~\cite{Tim20} (and references therein). Prediction of nuclear properties based on a realistic description of the strong interaction is at the heart of the ab initio effort in low-energy nuclear theory. Ab initio calculations have long been limited to light nuclei~\cite{Navratil16}, but with the ever-increasing computing power and its associated decreasing cost, ab initio calculations for many more nuclei are now in development~\cite{Chipo15}. These approaches are now used not only for predictions of binding energies but also to calculate one nucleon overlap functions~\cite{{Navratil04},{Fla13}} and nucleon optical potentials~\cite{{Rot17},{Rot18}}. Developments of optical potentials calculations using microscopic models have also been recently undertaken~\cite{Dick19} and the most recent WLH~\footnote{Whitehead, Lim and Holt} microscopic global optical potential could be very useful for the future transfer reaction experiments involving proton and neutron-rich isotopes~\cite{White20}. However, as Timofeyuk and Johnson pointed out so well \textit{"providing an input from ab-initio approaches to a transfer reaction amplitude based on an oversimplified distorted-wave approximation does not make the reaction description truly microscopic. To date only four truly ab-initio calculations of one-nucleon transfer have been published"}~\cite{Tim20}, involving light nuclei not heavier than $^8$Li~\cite{{Arai11},{Navr12},{Delt14},{Raim16}}.

Despite their use since more than 50 years, transfer reactions remain a powerful method in nuclear astrophysics which is still promised to have a bright future in the forthcoming decades to provide a better insight on the reactions that govern the Cosmos.

\section*{Acknowledgments}
NdS thanks Anne Meyer for providing angular distributions and $\alpha$-particle partial widths in case of $Q=8$ for the \clt\ study. NdS also thanks Sarah Harrouz for making available material concerning her $^{30}$Si($^3$He,d)$^{31}$P analysis prior to publication. FH and NdS thank D. Beaumel for a careful reading of the manuscript and stimulating discussions.

\bibliographystyle{frontiersinHLTH&FPHY} 
\bibliography{Transfer}
\end{document}